\documentclass[aps, prb, twocolumn,superscriptaddress,longbibliography,amsmath,amssymb]{revtex4-2}
\usepackage{graphicx}
\usepackage{dcolumn}
\usepackage{bm}
\usepackage{multirow,array}
\usepackage[pdftex,colorlinks=true,hypertexnames=false]{hyperref}

\usepackage{booktabs}

\begin{document}

\title{Sb-5$p$-mediated antiparallel Nd–Fe coupling in Nd$_3$Fe$_3$Sb$_7$}

\author{S. Palazzese}
\affiliation{Dresden High Magnetic Field Laboratory (HLD-EMFL) and Würzburg-Dresden Cluster of Excellence ctd.qmat, Helmholtz-Zentrum Dresden-Rossendorf, 01328 Dresden, Germany}%
\affiliation{Institute of Solid State and Materials Physics, Technische Universität Dresden, 01062 Dresden, Germany}
\author{Sh. Yamamoto}
\email{s.yamamoto@hzdr.de}
\affiliation{Dresden High Magnetic Field Laboratory (HLD-EMFL) and Würzburg-Dresden Cluster of Excellence ctd.qmat, Helmholtz-Zentrum Dresden-Rossendorf, 01328 Dresden, Germany}%
\author{T. Kotte}
\affiliation{Dresden High Magnetic Field Laboratory (HLD-EMFL) and Würzburg-Dresden Cluster of Excellence ctd.qmat, Helmholtz-Zentrum Dresden-Rossendorf, 01328 Dresden, Germany}%
\author{R. Ray}
\affiliation{Leibniz IFW Dresden, Helmholtzstr. 20, 01069 Dresden, Germany}
\affiliation{Department of Physics, Birla Institute of Technology Mesra, Ranchi 835215, India}%
\author{F. Pabst}
\affiliation{Faculty of Chemistry and Food Chemistry and Würzburg-Dresden Cluster of Excellence ctd.qmat, Technische Universität Dresden, 01062 Dresden, Germany}
\author{S. Chattopadhyay}
\affiliation{Dresden High Magnetic Field Laboratory (HLD-EMFL) and Würzburg-Dresden Cluster of Excellence ctd.qmat, Helmholtz-Zentrum Dresden-Rossendorf, 01328 Dresden, Germany}
\affiliation{UGC-DAE Consortium for Scientific Research Mumbai Centre, 246-C CFB, BARC Campus, Mumbai 400085, India}
\author{D. I. Gorbunov}
\affiliation{Dresden High Magnetic Field Laboratory (HLD-EMFL) and Würzburg-Dresden Cluster of Excellence ctd.qmat, Helmholtz-Zentrum Dresden-Rossendorf, 01328 Dresden, Germany}%
\author{T. Herrmannsdörfer}
\affiliation{Dresden High Magnetic Field Laboratory (HLD-EMFL) and Würzburg-Dresden Cluster of Excellence ctd.qmat, Helmholtz-Zentrum Dresden-Rossendorf, 01328 Dresden, Germany}%
\author{K. Kummer}
\affiliation{European Synchrotron Radiation Facility, 38043 Grenoble Cedex 9, France}%
\author{A. Kurtanidze}
\affiliation{Dresden High Magnetic Field Laboratory (HLD-EMFL) and Würzburg-Dresden Cluster of Excellence ctd.qmat, Helmholtz-Zentrum Dresden-Rossendorf, 01328 Dresden, Germany}
\affiliation{Institute of Solid State and Materials Physics, Technische Universität Dresden, 01062 Dresden, Germany}
\author{E. Weschke}
\affiliation{Helmholtz-Zentrum Berlin für Materialien und Energie, BESSY II, 12489 Berlin, Germany
}%
\author{O. Prokhnenko}
\affiliation{Helmholtz-Zentrum Berlin für Materialien und Energie, BESSY II, 12489 Berlin, Germany
}%
\author{K. Yamamoto}
\affiliation{Institute for Molecular Science, Okazaki, Aichi 444-8585, Japan}
\author{Y. Kotani}
\affiliation{Japan Synchrotron Radiation Research Institute, SPring-8, Sayo, Hyogo 679-5198, Japan}
\author{H. Nojiri}
\affiliation{Institute for Materials Research, Tohoku University, Sendai, 980-8577, Japan}%
\author{M. Ruck}
\affiliation{Faculty of Chemistry and Food Chemistry and Würzburg-Dresden Cluster of Excellence ctd.qmat, Technische Universität Dresden, 01062 Dresden, Germany}
\author{Manuel Richter}
\affiliation{Leibniz IFW Dresden, Helmholtzstr. 20, 01069 Dresden, Germany}
\affiliation{Dresden Center for Computational Materials Science (DCMS), TU Dresden, 01062 Dresden, Germany}
\author{J. Wosnitza}
\affiliation{Dresden High Magnetic Field Laboratory (HLD-EMFL) and Würzburg-Dresden Cluster of Excellence ctd.qmat, Helmholtz-Zentrum Dresden-Rossendorf, 01328 Dresden, Germany}%
\affiliation{Institute of Solid State and Materials Physics, Technische Universität Dresden, 01062 Dresden, Germany}%

\date{\today}

\begin{abstract}

We uncover the microscopic mechanism underlying the antiparallel Nd–Fe coupling in the intermetallic 4$f$–3$d$–5$p$ compound Nd$_3$Fe$_3$Sb$_7$ by means of high-field x-ray magnetic circular dichroism (XMCD) spectroscopy, complemented by magnetization measurements and density-functional-theory (DFT) calculations. This coupling is unexpected for intermetallics composed of light rare-earth and transition-metal elements. Sb $M_{4,5}$-edge XMCD measurements indicate a clear spin polarization of Sb 5$p$ states antiparallel to the Fe 3$d$ moments. DFT calculations reveal the site-dependent character of the Sb moments and, together with the Fe–Sb antiparallel alignment seen by XMCD, identify a Fe 3$d$–Sb 5$p$–Nd 5$d$ hybridization pathway for the antiparallel Nd–Fe coupling. Pulsed-field XMCD reveals a field-induced spin reorientation through the evolution of the $c$-axis projections of the Nd and Fe moments. Nd$_3$Fe$_3$Sb$_7$, thus, provides a platform for exploring antiparallel rare-earth--transition-metal ($R$--$T$) coupling mediated by transition-metal 3$d$–ligand $p$–rare-earth 5$d$ hybridization in light-$R$--$T$ intermetallics with $p$-elements.

\end{abstract}

\maketitle

The incorporation of pnictogens into rare-earth ($R$)–transition-metal ($T$) compounds has led to the discovery of numerous materials with unconventional magnetism and superconductivity \cite{de2008magnetic,wu2008theory,yin2011kinetic,fernandes2022iron,chen2016complex,han2013superconductivity}. Antimony-based systems, in particular, crystallize intricate Sb networks that can influence both magnetic interactions and electronic structure \cite{bud1998anisotropic,chan2000synthesis,mills2002chains,brown2006efficiency}. Such structural complexity provides fertile ground for realizing novel exchange mechanisms in $R$–$T$ intermetallics, where the magnetic coupling is sensitively governed by the nature of chemical bonding and electronic hybridization.

\begin{figure*}[t]
	\begin{center}
		\includegraphics[width=0.75\linewidth]{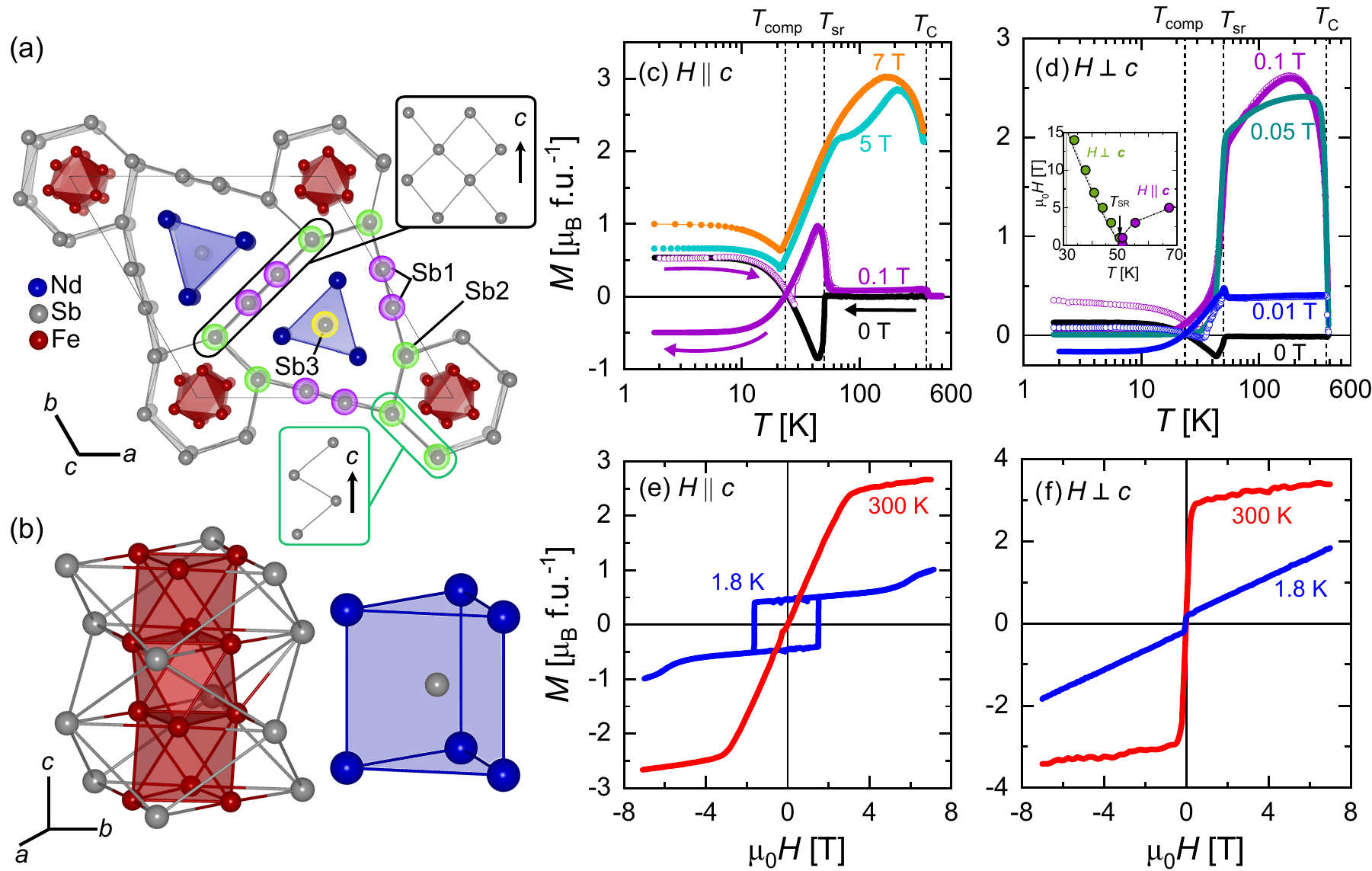}
	\end{center}
	\caption{(a), (b) Crystal structure of Nd$_3$Fe$_3$Sb$_7$ and Fe/Nd sublattices embedded in the Sb framework. Temperature-dependent magnetization for (c) $H$ $\parallel$ $c$ and (d) $H$ $\perp$ $c$; open and filled symbols denote data taken during heating after zero-field cooling and during field cooling, respectively, except for the zero-field data (black), which were obtained upon cooling in zero field. Inset: $H$-$T$ diagram of the spin-reorientation transition for $H$ $\perp$ $c$ (green) and $H$ $\parallel$ $c$ (purple). Field-dependent magnetization at 1.8 K (blue) and 300 K (red) for (e) $H$ $\parallel$ $c$ and (f) $H$ $\perp$ $c$.}
	\label{fig:1}
\end{figure*}

In $R$–$T$ intermetallic compounds, magnetic coupling between rare-earth and transition-metal sublattices often depends on the position of the $R$ element in the lanthanide series. Light rare-earth elements tend to couple ferromagnetically with $T$ moments, while heavy rare-earth elements more commonly display ferrimagnetic or antiferromagnetic alignment \cite{franse1993magnetic,campbell1972indirect,nguyen2024ligand}, provided that the 4$f$ electrons remain well localized. Only a few light–$R$ intermetallics exhibit antiparallel $R$–$T$ coupling or compensation-like behavior, including the cobalt pnictides $R_2$Co$_{12}X_7$ ($R$ = Pr, Nd; $X$ = P, As) and $R$Co$_2$As$_2$ ($R$ = Pr, Nd) \cite{ohta2016anomalous,tan2016synthesis,thompson2014synthesis}, as well as $R$CrSb$_3$ ($R$ = Pr, Nd) antimonides \cite{jackson2006effect,inamdar2008anisotropic,inamdar2009interplay,inamdar2009anisotropic}. For $R$CrSb$_3$, the $R$–Cr coupling has been discussed in terms of 3$d$–4$f$ interactions, RKKY-type exchange, and conduction-electron polarization. We recently uncovered antiparallel Pr–Fe alignment in Pr$_3$Fe$_3$Sb$_7$ \cite{pabst2022}. In that study, we discussed that the unusual magnetic-moment alignment may be influenced by the structural framework of Sb atoms and the internal field of the Fe sublattice. We proposed that enhanced hybridization between Fe 3$d$ and rare-earth 5$d$ states, possibly facilitated by the proximity of Sb, could underlie this behavior. Nevertheless, across these reported light-$R$ intermetallics, including Pr$_3$Fe$_3$Sb$_7$, the microscopic origin of the antiparallel $R$–$T$ coupling and, in particular, the role of spin-polarized ligand $p$ states, has remained unresolved.

Related compensation, spin-reorientation, and magnetization-reversal phenomena are also known in rare-earth perovskites and double perovskites \cite{toulemonde1998orientation,munoz2000magnetic,kumar2025exchange,pal2019peculiar}. In these oxides, the magnetic behavior is generally discussed in terms of oxygen-mediated transition-metal exchange interactions, such as superexchange or double exchange, together with the exchange field acting on the rare-earth moments and magnetic anisotropy. These studies highlight the exceptional variety of rare-earth–transition-metal coupling across different material classes, but they do not establish a spin-polarized $p$-block ligand state as the microscopic pathway mediating antiparallel light-rare-earth–transition-metal coupling in intermetallics. Here, we identify a novel indirect coupling mechanism mediated by polarized Sb 5$p$ states.

Building on these insights, we turn to the isostructural compound Nd$_3$Fe$_3$Sb$_7$ to elucidate the microscopic origin of possible antiparallel light $R$-$T$ coupling. This compound crystallizes in a hexagonal structure (space group $P6_3/m$) \cite{nasir2010} composed of alternating face-sharing Fe octahedra at the corners of the unit cell and two triangular stacks of Nd atoms, all embedded within a three-dimensional Sb framework [Figs. \ref{fig:1}(a) and \ref{fig:1}(b)]. The resulting architecture features columnar arrangements of Nd and Fe atoms along the $c$ axis, spatially separated by the Sb sublattice, making Nd$_3$Fe$_3$Sb$_7$ a promising candidate for investigating Sb-mediated magnetic interactions.

To clarify the magnetic coupling mechanisms in this system, we employed element-specific x-ray magnetic circular dichroism (XMCD) at very high magnetic fields to resolve the individual spin orientations and magnetic contributions of the Nd, Fe, and Sb atoms. Our measurements reveal antiparallel alignment between the Nd and Fe moments, along with a finite XMCD signal from Sb, indicating that the Sb 5$p$ states become spin polarized. Our experimental data and DFT-based calculations support a two-fold antiparallel spin-coupling pathway, Fe(3$d$)–Sb(5$p$)–Nd(5$d$), which yields the antiparallel Fe–Nd alignment. This metallic, hybridization-driven Sb-mediated exchange is distinct from conventional indirect-exchange mechanisms, such as superexchange \cite{imada1998metal} in localized-moment systems and double exchange \cite{salamon2001physics} in mixed-valence systems.

We synthesized single crystals of Nd$_3$Fe$_3$Sb$_7$ using an inert, molten Bi flux as reaction medium \cite{NdFeSbSupplemental}. We carried out magnetization measurements using SQUID and vibrating-sample magnetometry (VSM) in magnetic fields up to 14 T. For each temperature-dependent magnetization measurement, we heated the sample above the Curie temperature, $T_\mathrm{C}$, prior to cooling, in order to ensure consistent magnetic initialization. To gain element-specific insight into the magnetic properties, we performed x-ray absorption spectroscopy (XAS) and x-ray magnetic circular dichroism (XMCD) measurements at the Nd \textit{M}$_{4,5}$ (3\textit{d} $\rightarrow$ 4\textit{f}), Fe \textit{L}$_{2,3}$ (2\textit{p} $\rightarrow$ 3\textit{d}), and Sb \textit{M}$_{4,5}$ (3\textit{d} $\rightarrow$ 5\textit{p}) absorption edges, which are sensitive to Nd 4$f$, Fe 3$d$, and Sb 5$p$ electrons, respectively. We performed static-field measurements at the ID32 beamline of the ESRF \cite{brookes2018beamline,kummer2016high} and the BL25SU beamline of SPring-8 \cite{senba2016upgrade,nakamura2005development}, and pulsed-field measurements up to 30 T at the UE46-PGM1 beamline of BESSY II \cite{weschke2018ue46,yamamoto2024field}. We performed all XMCD measurements with the magnetic field applied along the $c$ axis and detected by total-electron yield (TEY).

Temperature- and field-dependent magnetization data by means of SQUID magnetometry for Nd$_3$Fe$_3$Sb$_7$ [Figs. \ref{fig:1}(c)–\ref{fig:1}(f)] reveal a magnetic transition at $T_\mathrm{C}$ $\sim$ 390 K, followed by a spin-reorientation transition (SRT) at $T_\mathrm{sr}$ $\sim$ 50 K and a compensation point at $T_\mathrm{comp} \sim 23.5$ K, observed upon cooling in zero field. This behavior is observed for both $H$ $\parallel$ $c$ and $H$ $\perp$ $c$, indicating ferrimagnetic order of the Fe and Nd sublattices. The occurrence of both a spin-reorientation transition and a compensation point, together with the change from predominantly in-plane magnetization above $T_\mathrm{sr}$ to finite magnetization components both parallel and perpendicular to $c$ below $T_\mathrm{sr}$, closely resembles the bulk-magnetization behavior reported for the isostructural compound Pr$_3$Fe$_3$Sb$_7$ \cite{pabst2022}. The anisotropic magnetization responses for $H \parallel c$ and $H \perp c$ demonstrate strong magnetic anisotropy, whose effects on the negative low-field magnetization and the field dependence of $T_\mathrm{sr}$ are discussed in the Supplemental Material \cite{NdFeSbSupplemental}. 

At 1.8 K, the presence of remanent magnetizations of 0.48 $\mu_\mathrm{B}$/f.u. for $H$ $\parallel$ $c$ and 0.13 $\mu_\mathrm{B}$/f.u. for $H$ $\perp$ $c$ [Figs. \ref{fig:1}(e) and \ref{fig:1}(f)] indicates that the net magnetization is not confined strictly to either the $c$ axis or the $ab$ plane. In addition, our specific-heat data show field-induced transitions near $T_\mathrm{comp}$ \cite{NdFeSbSupplemental}, which implies field-induced spin reorientation of the antiferromagnetically coupled sublattices. The associated entropy change $\Delta S$, obtained by integrating $C/T$ over the measured temperature window, remains below 0.5$R$ at fields up to 19 T \cite{NdFeSbSupplemental}. This modest entropy release indicates that the field-induced transitions near $T_\mathrm{comp}$ involve only the limited entropy associated with the spin-reorientation process, rather than a substantial rearrangement of the Nd or Fe magnetic sublattices. A fully saturated collinear state is not achieved even in pulsed fields up to 58 T \cite{NdFeSbSupplemental}, highlighting the robustness of the antiferromagnetic Nd–Fe intersublattice coupling.

The observed compensation behavior indicates ferrimagnetic order. An antiparallel $R$–$T$ total-moment alignment is rare for light rare-earth elements \cite{pabst2022,ohta2016anomalous,tan2016synthesis,thompson2014synthesis}, for which the rare-earth total magnetic moment typically aligns parallel to the transition-metal moment.  In this context, the antiparallel coupling between the Nd total moment and the Fe moment in a light-$R$–$T$ intermetallic indicates an exchange pathway beyond the conventional light-$R$–$T$ coupling picture, potentially involving Sb-mediated interactions or hybridization effects. To clarify the microscopic origin of these interactions, we now turn to our element-resolved XMCD results.

\begin{figure}[t]
	\begin{center}
		\includegraphics[width=0.99\columnwidth]{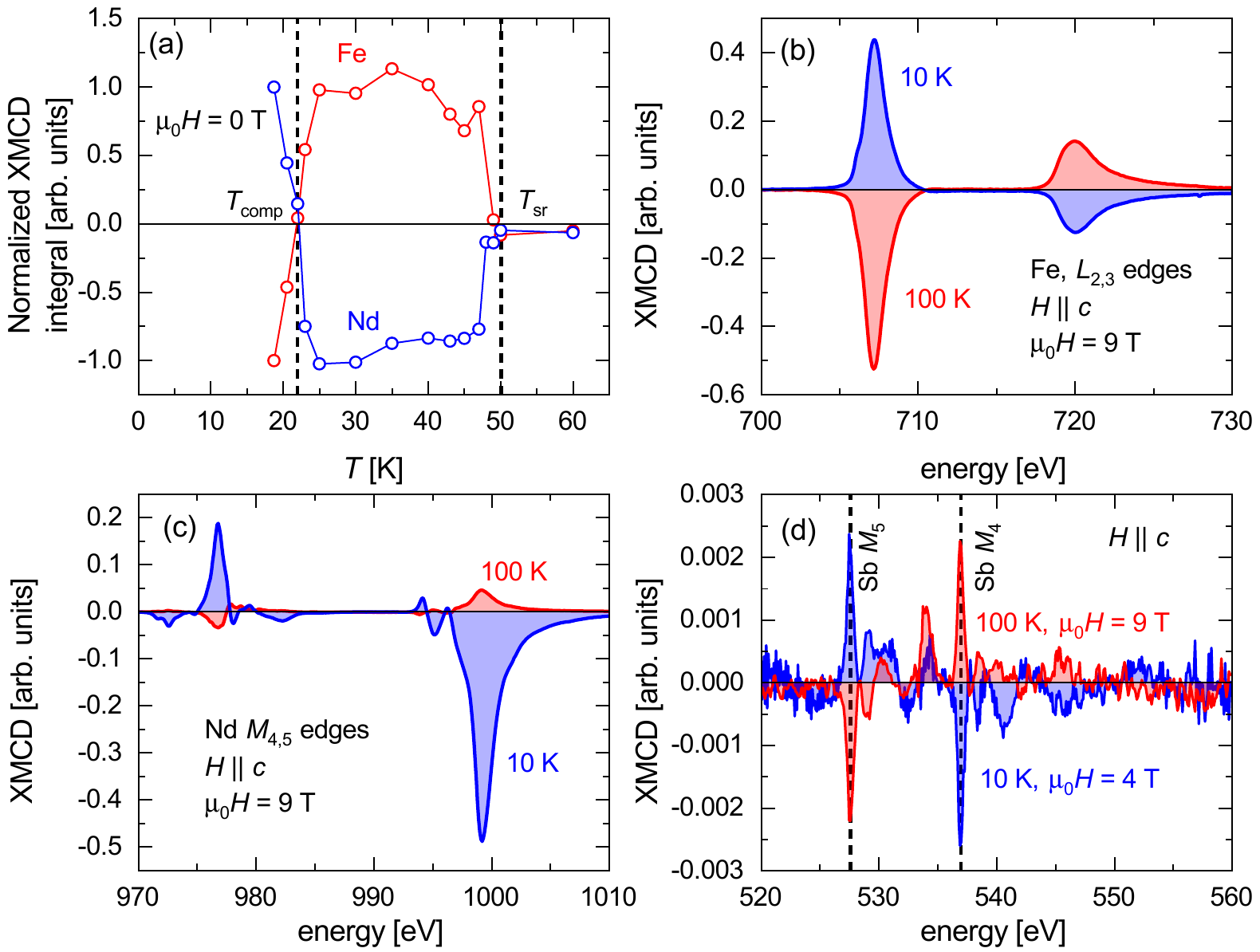}
	\end{center}
	\caption{(a) Temperature dependence of Nd $M_4$- and Fe $L_3$-edge XMCD integrals measured at each temperature in remanence after a 30 T field pulse. XMCD spectra at the (b) Fe $L_{2,3}$ and (c) Nd $M_{4,5}$ edges at 10 and 100 K. (d) Sb $M_{4,5}$-edge XMCD spectra.}
	\label{fig:2}
\end{figure}

Figure \ref{fig:2}(a) shows the temperature dependence of the XMCD integrals at the Nd $M_4$ and Fe $L_3$ edges, measured at each temperature in remanence after applying a 30 T pulsed field along the $c$ axis. The integrals are normalized to their values at the lowest measured temperature (19 K) and plotted with reversed sign relative to Figs. \ref{fig:2}(b) and \ref{fig:2}(c), so that positive (negative) values correspond to $c$-axis moment projections aligned parallel (antiparallel) to the direction of the preceding field pulse. Above 50 K, the remanent Nd and Fe XMCD integrals are nearly zero. Since the measurements [Fig. \ref{fig:2}(a)] probe only the remanent $c$-axis projections, these nearly vanishing signals are consistent with the predominantly in-plane magnetic response above $T_\mathrm{sr}$ observed by bulk magnetization. Below 50 K, finite remanent $c$-axis projections of both Nd and Fe emerge with clear antiparallel alignment, consistent with the SRT at $T_\mathrm{sr}$ $\sim$ 50 K seen via magnetometry (Fig. \ref{fig:1}). The sign change of both the Nd and Fe XMCD signals occurs near 22 K, close to the macroscopic compensation temperature. At a ferrimagnetic compensation point, the local Nd and Fe moments need not vanish. Instead, their contributions to the net magnetization compensate each other. Under the present remanent XMCD condition, where a 30 T field pulse was applied before the measurement at each temperature, the vanishing XMCD signals can most naturally arise from cancellation of the $c$-axis projections averaged over oppositely oriented ferrimagnetic domains, without requiring an additional spin reorientation. Alternatively, the local $c$-axis projections may be reduced by an additional spin reorientation toward predominantly in-plane moment orientations close to $T_\mathrm{comp}$ or by a decoupled, paramagnetic-like sublattice response near compensation \cite{boada2010decoupling}. As TEY-mode XMCD probes the surface region to a depth of a few nanometers, its surface sensitivity likely explains the slight offset between the bulk compensation temperature, determined by magnetometry, $T_\mathrm{comp}$ = 23.5 K, and the sign-change temperature obtained from soft XMCD measurements, 22 K.

Figures \ref{fig:2}(b) and \ref{fig:2}(c) show the XMCD spectra at 10 and 100 K at the Fe $L_{2,3}$ and Nd $M_{4,5}$ edges, respectively. The corresponding XAS spectra are provided in the Supplemental Material \cite{NdFeSbSupplemental}. Even at a magnetic field of 9 T, the XMCD signals at the Fe $L_3$ and Nd $M_4$ edges exhibit opposite signs at both temperatures, indicating that their $c$-axis components are oppositely oriented.

In addition, the XMCD spectra at the Sb $M_{4,5}$ edges show two main near-edge peaks at both 10 and 100 K [Fig. \ref{fig:2}(d)], providing evidence for induced magnetic moments of the Sb 5$p$ electrons, as discussed in the Supplemental Material \cite{NdFeSbSupplemental}. The nearly equal magnitudes of the XMCD signals at the Sb $M_4$ and $M_5$ edges indicate that the induced magnetism in Sb arises predominantly from spin moments, with negligible orbital contributions. Furthermore, the XMCD amplitudes at the Fe $L_3$ edge (2$p \rightarrow 3d$ transition) and the Sb $M_5$ edge (3$d \rightarrow 5p$ transition) have the same sign. Because the relationship between the XMCD sign and the direction of the spin moment is opposite for $p \rightarrow d$ and $d \rightarrow p$ transitions \cite{ye2015carrier,thole1992x,carra1993x}, the same XMCD sign at these two edges evidences antiparallel coupling between the Fe 3$d$ and Sb 5$p$ spin moments. The relative orientation of the magnetic moments of the Fe 3$d$, Nd 4$f$, and Sb 5$p$ orbitals is as follows (↑ and ↓ denote directions parallel and antiparallel to the applied magnetic field, respectively): Fe↓ – Sb↑ – Nd↑ at 10 K, and Fe↑ – Sb↓ – Nd↓ at 100 K.

\begin{table}[t]
\caption{\label{tab1}
Spin ($m_s$), orbital ($m_l$), and total ($m_{tot}$) magnetic moments (in $\mu_\mathrm{B}$/atom) of Fe 3$d$ and Nd 4$f$ electrons, determined by XMCD measurements in 9 T at 10 and 100 K. For Fe, $m^{\mathrm{eff}}_s$ (see text) and $m^{\mathrm{eff}}_{tot}$ (= $m^{\mathrm{eff}}_s$ + $m_l$) refer to effective moments. The experimental values are rounded to two decimal places, while the total moments were calculated from the unrounded spin and orbital moments. DFT-calculated total moments for Nd 5$d$, Fe 3$d$, and Sb 5$p$ electrons are shown in the lower section. The largest possible Nd 4$f$ moment is defined by the $z$ component of the $L\cdot S$-atomic value; this value is strongly reduced by crystal field effects in a chemical environment. Numbers in parentheses indicate the site degeneracy for inequivalent Sb sites, and $\langle m_\mathrm{Sb}\rangle$ represents the average moment over all Sb sites.
}
\vspace{0.5em}
\begin{center}
{\setlength{\tabcolsep}{0pt}%
\begin{tabular}{@{}l@{}}
\toprule
\begin{tabular*}{\linewidth}{@{\extracolsep{\fill}} l ccc ccc}
\multicolumn{7}{c}{Experiment} \\
\multirow{2}{*}{$T$ (K)} & \multicolumn{3}{c}{Nd} & \multicolumn{3}{c}{Fe} \\
\cmidrule(r){2-4}\cmidrule(l){5-7}
 & $m_s$ & $m_l$ & $m_{\mathrm{tot}}$ & $m^{\mathrm{eff}}_s$ & $m_l$ & $m^{\mathrm{eff}}_{\mathrm{tot}}$ \\
 \cmidrule[0.25pt]{1-7}
10  & $-1.28$ & $2.98$ & $1.70$ & $-1.17$ & $-0.07$ & $-1.25$ \\
100 & $0.13$ & $-0.28$ & $-0.15$ & $1.24$ & $0.15$ & $1.40$ \\
\end{tabular*}
\\[-0.25\baselineskip]
\midrule
\begin{tabular*}{\linewidth}{@{\extracolsep{\fill}} l ccccccc}
 & \multicolumn{1}{c}{$\langle L \cdot S \rangle_z$}
 & \multicolumn{6}{c}{DFT calculation} \\
\cmidrule(r){2-2}\cmidrule(l){3-8}
$T$ (K) & $m^{{\mathrm{Nd}\text{-}4}f}_{\mathrm{tot}}$
& $m^{{\mathrm{Nd}\text{-}5}d}_{\mathrm{tot}}$
& $m^{{\mathrm{Fe}\text{-}3}d}_{\mathrm{tot}}$
& Sb1 (6) & Sb2 (6) & Sb3 (2)
& $\langle m_{\mathrm{Sb}}\rangle$ \\
\cmidrule[0.25pt]{1-8}
0 & 3.27 & -0.16 & -2.57 & 0.03 & 0.20 & 0.13 & 0.12 \\
\end{tabular*}
\\[-0.25\baselineskip]
\bottomrule
\end{tabular}%
}
\end{center}

\end{table}

\begin{figure}[t]
	\begin{center}
		\includegraphics[width=0.99\columnwidth]{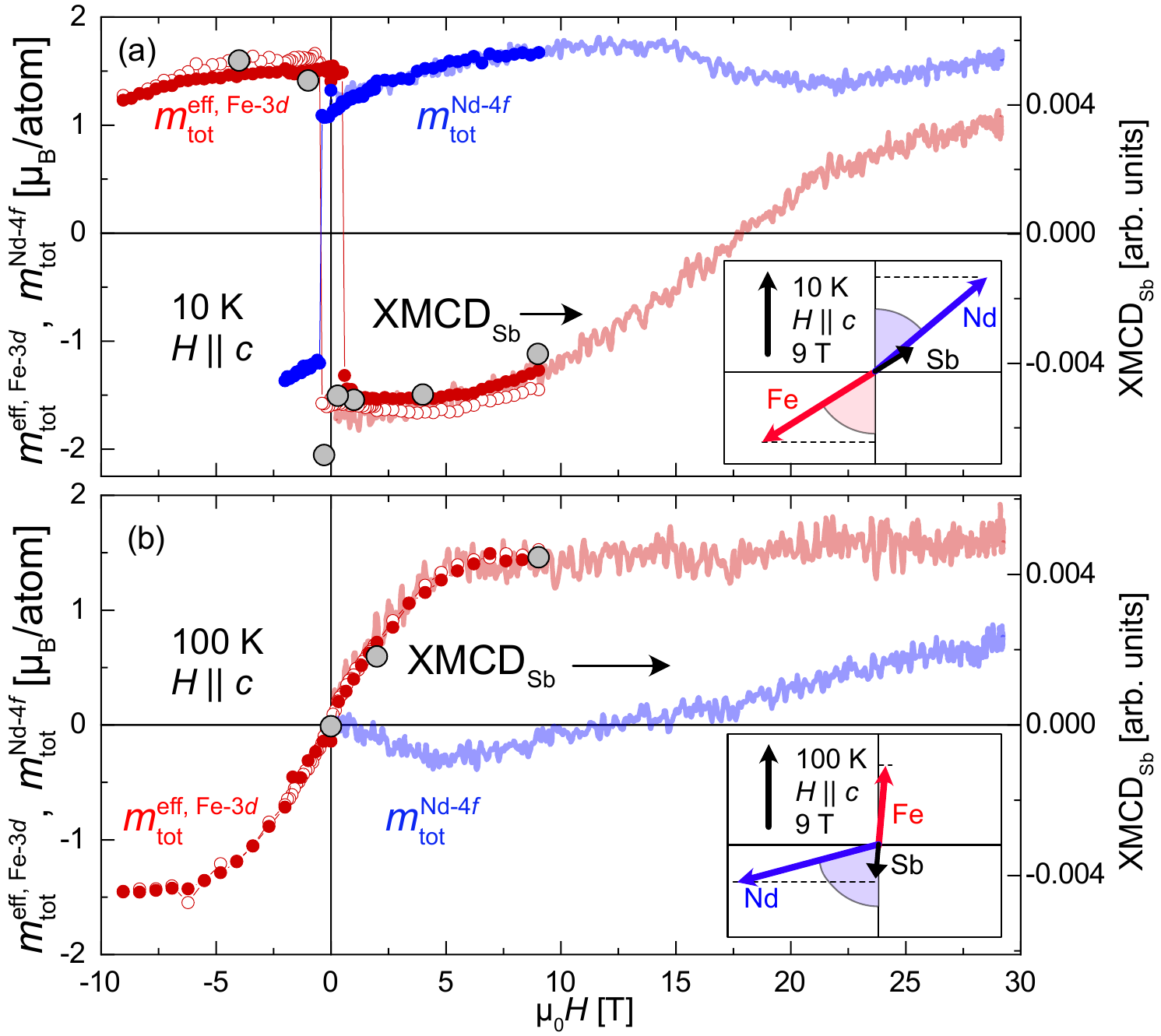}
	\end{center}
	\caption{Field dependence of $c$-axis projected effective Fe 3$d$ total magnetic moments ($m^{\mathrm{eff, Fe}\text{-}3d}_{\mathrm{tot}}$, red), Nd 4$f$ total magnetic moments ($m^{\mathrm{Nd}\text{-}4f}_{\mathrm{tot}}$, blue), and Sb XMCD defined as XMCD($M_4$) $-$ XMCD($M_5$) (gray circles) at (a) 10 K and (b) 100 K. Filled and open circles denote static-field data taken on the up- and down-sweeps, respectively. Light solid lines are data taken in pulsed fields for Fe and Nd. Insets depict schematic possible noncollinear magnetic configurations consistent with the $c$-axis projected moments at 9 T.}
	\label{fig:3}
\end{figure}

We extracted the Fe and Nd moments at 10 and 100 K from the XMCD data via sum rules (Table \ref{tab1}). The Fe spin moment effectively includes the contribution from the magnetic dipole moment ($m^{\mathrm{eff}}_s$ = $m_s-7\left\langle T_z \right\rangle$, where $\left\langle T_z \right\rangle$ is the expectation value of the magnetic dipole operator). Details of the analysis are provided in the Supplemental Material \cite{NdFeSbSupplemental}. The total magnetic moments ($m_{tot}$) of Nd and Fe exhibit opposite signs at both 10 and 100 K, confirming the antiparallel alignment inferred from the XMCD spectra. The experimentally determined $m_{tot}$ values from the XMCD sum rules are smaller than those obtained from density-functional-theory (DFT) calculations assuming a collinear magnetic structure \cite{NdFeSbSupplemental}. Since XMCD measures the moment component projected along the x-ray propagation direction, i.e., along the crystallographic $c$ axis, this reduction may indicate that the Nd and Fe moments are not fully aligned along $c$ at 9 T. This interpretation is consistent with the low-temperature magnetization data [Figs. \ref{fig:1}(e) and \ref{fig:1}(f)], where $M$($H$) for $H \parallel c$ is not saturated at 7 T, indicating that the moments are not fully polarized along $c$ at 7 T. The finite remanent magnetization for $H \perp c$ further suggests that the low-temperature magnetic state is not described by a simple collinear configuration along the $c$ axis. This is also in line with a magnetic structure involving canting away from the $c$ axis proposed for the isostructural compound Pr$_3$Fe$_3$Sb$_7$ \cite{pabst2022}.

The DFT results further indicate that the largest spin polarization occurs at the Sb2 site, which is located closest to Fe and situated between the Nd and Fe atoms [Fig. \ref{fig:1}(a)]. This positions them suitably for the distinctive role of mediating the exchange coupling between the Fe and Nd sublattices.

To investigate the field dependence of the magnetic coupling between Fe and Nd, we performed pulsed-field XMCD measurements up to 30 T. Figures \ref{fig:3}(a) and \ref{fig:3}(b) show the element-specific magnetic moments projected along the $c$ axis as a function of field at 10 and 100 K, together with static-field results up to 9 T. The Nd and Fe data show finite remanent magnetization and a finite coercive field at 10 K, and their absence at 100 K, consistent with the spin reorientation near $T_\mathrm{sr}$ $\sim$ 50 K. At 10 K, the projected Nd moment increases up to $\sim$ 10 T, while at the same time the projected Fe moment continuously changes towards the direction of the applied field. But $M_\mathrm{Nd}$ subsequently decreases slightly before rising again above 25 T, indicating a nonmonotonic response. This behavior reflects a competition between Zeeman interaction and antiferromagnetic Nd–Fe intersublattice exchange interaction, with the latter dominating at intermediate fields. At 100 K, the $c$-axis projection of the Fe moment shows only a weak increase above 5 T, while that of the Nd 4$f$ moment changes sign with field. More generally, at both 10 and 100 K, the projected Nd and Fe moments evolve continuously with field, without a distinct transition field. Together with the absence of saturation in pulsed-field magnetization measurements up to 58 T and the reduced XMCD-derived moment magnitudes compared with the collinear DFT values, these results are consistent with a continuous field-driven reorientation of the Nd and Fe moments in a noncollinear magnetic configuration.

The field dependence of the Sb XMCD amplitudes, obtained from static-field measurements, closely tracks that of $M_\mathrm{Fe}$, indicating that the Sb 5$p$ moments are not directly responsive to the external field but are instead induced by hybridization with Fe 3$d$ states. The same XMCD signs of Fe $L_3$ and Sb $M_5$, along with the excellent scaling between $M_\mathrm{Fe}$ and $\mathrm{XMCD}_\mathrm{Sb}$, support a picture, in which the Sb moments are antiparallel to and driven by the Fe sublattice. This type of antiparallel $p$–$d$ spin polarization, arising from antibonding hybridization between spin-polarized $d$ and ligand $p$ orbitals, has been reported in intermetallics \cite{kimura1997magnetic, holm2002xmcd} and in $p$-ligand systems including oxides, chalcogenides, and topological magnets \cite{murakawa2010ferroelectricity, yamagami2021itinerant, ye2015carrier, wang2024double}, but has not previously been observed in light rare-earth intermetallics. 

\begin{figure}[t]
	\begin{center}
		\includegraphics[width=0.9\linewidth]{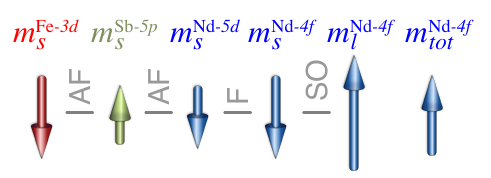}
	\end{center}
	\caption{Schematic exchange pathways in Nd$_3$Fe$_3$Sb$_7$. Arrows indicate local moments projected onto the crystallographic $c$ axis at low temperatures. AF and F indicate antiferromagnetic and ferromagnetic coupling, respectively. SO denotes spin–orbit coupling, which leads to antiparallel alignment between the Nd 4$f$ spin and orbital moments.
}
	\label{fig:4}
\end{figure}

The Sb 5$p$ spin polarization appears to be induced by Fe. It is, thus, passive in the magnetic order but closely follows the Fe magnetization. Since the Sb2 atoms structurally bridge the Fe and Nd sublattices, Sb mediates an indirect exchange between the two. The Sb 5$p$ band is centered at higher energy than the Fe 3$d$ band. Hence, Sb 5$p$ orbitals adjacent to Fe couple antiferromagnetically with Fe 3$d$ (antiparallel spin exchange coupling due to antibonding hybridization). The same argument can be repeated for the coupling between Sb 5$p$ and adjacent Nd 5$d$ orbitals, which are yet less populated (i.e., higher in energy) than Sb 5$p$. The resulting coupling between Fe 3$d$ and Nd 5$d$, via Sb 5$p$, is parallel. This contrasts with the conventional picture for light $R$–$T$ intermetallics, where the $T$-3$d$–$R$-5$d$ exchange is antiferromagnetic and leads to parallel alignment between the total $R$ moment and the $T$ moment. Finally, ferromagnetic intra-atomic exchange between Nd 5$d$ and Nd 4$f$ spins, together with the large antiparallel Nd 4$f$ orbital moment yields the observed ferrimagnetic alignment between the Nd total moment and the Fe moment (Fig. \ref{fig:4}). Consistent with this pathway, DFT yields an induced Nd 5$d$ moment of -0.16 $\mu_\mathrm{B}$/Nd, antiparallel to Sb 5$p$ and parallel to Fe 3$d$. Since Nd$_3$Fe$_3$Sb$_7$ is metallic, RKKY-type interactions are also expected to provide a minor contribution to the overall exchange network. Nevertheless, the element-selective XMCD data, the Sb-containing structural network, and the DFT spin polarization of Sb 5$p$ and Nd 5$d$ states provide direct support for a local Fe 3$d$–Sb 5$p$–Nd 5$d$ hybridization-driven pathway. We, therefore, identify this pathway as the microscopic mechanism underlying the antiparallel Nd–Fe coupling in Nd$_3$Fe$_3$Sb$_7$. 

We have presented detailed magnetization and XMCD results for Nd$_3$Fe$_3$Sb$_7$, establishing ferrimagnetic alignment between Nd and Fe sublattices with induced spin polarization on Sb. A finite XMCD signal at the Sb site indicates spin polarization antiparallel to Fe, consistent with $p$–$d$ hybridization. These findings are corroborated by our DFT calculations, which reproduce the observed antiparallel moments and reveal site-dependent Sb spin polarization. The combined signatures of field-induced reorientation, noncollinear magnetic configurations, and spin entropy highlight a complex magnetic energy landscape in Nd$_3$Fe$_3$Sb$_7$.
Together with the two-fold antiparallel spin coupling identified here, this complexity provides a novel platform for exploring ferrimagnetism and ligand spin polarization in light-rare-earth intermetallics and suggests broader relevance for other 3$d$–4$f$–$p$ intermetallics exhibiting unconventional magnetic alignment.

\begin{acknowledgments}
We acknowledge the support of the HLD at HZDR, a member of the European Magnetic Field Laboratory (EMFL), the Deutsche Forschungsgemeinschaft (DFG) through SFB 1143 (Project No.\ 247310070), and the W\"{u}rzburg-Dresden Cluster of Excellence on Complexity, Topology and Dynamics in Quantum Matter--$ctd.qmat$ (EXC 2147, Project No.\ 390858490). We acknowledge the European Synchrotron Radiation Facility (ESRF) for provision of synchrotron radiation facilities under proposal number HC-5356. Synchrotron radiation experiments were also performed at the UE46-PGM1 undulator beamline of BESSY II under the GIMRT/ICC-IMR collaboration, and at BL25SU of SPring-8 with the approval of the Japan Synchrotron Radiation Research Institute (JASRI) (Proposal No. 2021B1590 and 2021B2085). We gratefully acknowledge discussions with Ulrich Rößler, Jeroen van den Brink, and Helge Rosner. We also wish to express our sincere gratitude to Ulrike Nitzsche for maintaining the HPC facilities at IFW Dresden, and Sovam Maiti for maintaining the HPC facility at BIT Mesra, which were used for the calculations.

F.P. grew the single crystals. S.P., D.G., S.C., and T.H. carried out the bulk magnetization measurements. S.Y., K.K., A.K., E.W., O.P., K.Y., Y.K., and H.N. performed XMCD measurements. T.K. performed specific-heat measurements. S.P., S.Y., T.K., D.G., T.H., and K.K. analyzed the data. R.R. planned the DFT calculations with M.R., performed the DFT calculations, evaluated their results. All authors contributed to the interpretation of the data and to the writing and revising of the manuscript.

\end{acknowledgments}

\nocite{pabst2023synthetic,skourski2011high,kotte2022orbital,yasui2015temperature,chen1995experimental,kummer2013thin,jiang2013experimental,gurevich1999nexafs,teramura1996effect,fernandez2020influence,stoll2016magnetic,cid2017perpendicular,tcakaev2023intermixing,PBE1996,Koepernik1999,lejaeghere2016reproducibility,duffy2018imposing}
\bibliography{ref}

@article{de2008magnetic,
  author={de La Cruz, Clarina and Huang, Q and Lynn, JW and Li, Jiying and Ii, W Ratcliff and Zarestky, Jerel L and Mook, HA and Chen, GF and Luo, JL and Wang, NLa and Dai, Pengcheng},
  journal={Nature},
  title={Magnetic order close to superconductivity in the iron-based layered {L}a{O}$_{1-x}${F}$_x${F}e{A}s systems},
  year={2008},  
  number={7197}, 
  pages={899--902},
  volume={453},
  publisher={Nature Publishing Group UK London},
  doi={https://doi.org/10.1038/nature07057}
}

@article{wu2008theory,
  title={Theory of the magnetic moment in iron pnictides},
  author={Wu, Jiansheng and Phillips, Philip and Castro Neto, AH},
  journal={Phys. Rev. Lett.},
  volume={101},
  number={12},
  pages={126401},
  year={2008},
  publisher={APS},
  doi={https://doi.org/10.1103/PhysRevLett.101.126401}
}

@article{yin2011kinetic,
	title={Kinetic frustration and the nature of the magnetic and paramagnetic states in iron pnictides and iron chalcogenides},
	author={Yin, ZP and Haule, Kristjan and Kotliar, Gabriel},
	journal={Nat. Mater.},
	volume={10},
	number={12},
	pages={932--935},
	year={2011},
	publisher={Nature Publishing Group UK London},
	doi={https://doi.org/10.1038/nmat3120}
}

@article{fernandes2022iron,
	title={Iron pnictides and chalcogenides: a new paradigm for superconductivity},
	author={Fernandes, Rafael M and Coldea, Amalia I and Ding, Hong and Fisher, Ian R and Hirschfeld, PJ and Kotliar, Gabriel},
	journal={Nature},
	volume={601},
	number={7891},
	pages={35--44},
	year={2022},
	publisher={Nature Publishing Group UK London},
	doi={https://doi.org/10.1038/s41586-021-04073-2}
}

@article{chen2016complex,
	title={From complex magnetism ordering to simple ferromagnetism in two-dimensional {L}a{C}r{S}b$_3$ by hole doping},
	author={Chen, Haijie and Narayan, Awadhesh and Fang, Lei and Calta, Nicholas P and Shi, Fengyuan and Chung, Duck Young and Wagner, Lucas K and Kwok, Wai-Kwong and Kanatzidis, Mercouri G},
	journal={Phys. Rev. B},
	volume={94},
	number={13},
	pages={134411},
	year={2016},
	publisher={APS},
	doi = {https://doi.org/10.1103/PhysRevB.94.134411}
}

@article{han2013superconductivity,
	title={Superconductivity and strong intrinsic defects in {L}a{P}d$_{1-x}${B}i$_2$},
	author={Han, Fei and Malliakas, Christos D and Stoumpos, Constantinos C and Sturza, Mihai and Claus, Helmut and Chung, Duck Young and Kanatzidis, Mercouri G},
	journal={Phys. Rev. B},
	volume={88},
	number={14},
	pages={144511},
	year={2013},
	publisher={APS},
	doi={https://doi.org/10.1103/PhysRevB.88.144511}
}

@article{bud1998anisotropic,
	title={Anisotropic magnetic properties of light rare-earth diantimonides},
	author={Bud’ko, Sergey L and Canfield, PC and Mielke, CH and Lacerda, AH},
	journal={Phys. Rev. B},
	volume={57},
	number={21},
	pages={13624},
	year={1998},
	publisher={APS},
	doi = {https://doi.org/10.1103/PhysRevB.57.13624}
}

@article{chan2000synthesis,
	title={Synthesis, magnetic properties, and colossal magnetoresistance of {E}u$_{13.97}${G}d$_{0.03}${M}n{S}b$_{11}$},
	author={Chan, Julia Y and Kauzlarich, Susan M and Klavins, Peter and Liu, J-Z and Shelton, Robert N and Webb, David J},
	journal={Phys. Rev. B},
	volume={61},
	number={1},
	pages={459},
	year={2000},
	publisher={APS},
	doi = {https://doi.org/10.1103/PhysRevB.61.459}
}

@article{mills2002chains,
	title={Chains, planes, and antimonides},
	author={Mills, Allison M and Lam, Robert and Ferguson, Michael J and Deakin, Laura and Mar, Arthur},
	journal={Coord. Chem. Rev.},
	volume={233},
	pages={207--222},
	year={2002},
	publisher={Elsevier},
	doi = {https://doi.org/10.1016/S0010-8545(02)00097-8}
}

@article{brown2006efficiency,
	title={{Y}b14{M}n{S}b11: New high efficiency thermoelectric material for power generation},
	author={Brown, Shawna R and Kauzlarich, Susan M and Gascoin, Franck and Snyder, G Jeffrey},
	journal={Chem. Mater.},
	volume={18},
	number={7},
	pages={1873--1877},
	year={2006},
	publisher={ACS Publications},
	doi = {https://doi.org/10.1021/cm060261t}
}

@article{campbell1972indirect,
  title={Indirect exchange for rare earths in metals},
  author={Campbell, IA},
  journal={J. Phys. F: Met. Phys.},
  volume={2},
  number={3},
  pages={L47},
  year={1972},
  publisher={IOP Publishing},
  doi={10.1088/0305-4608/2/3/004}
}

@article{pabst2022,
	author = {Pabst, Falk and Palazzese, Sabrina and Seewald, Felix and Yamamoto, Shingo and Gorbunov, Denis and Chattopadhyay, Sumanta and Herrmannsd{\"{o}}rfer, Thomas and Ritter, Clemens and Finzel, Kati and Doert, Thomas and Klauss, Hans-Henning and Wosnitza, Jochen and Ruck, Michael},
	journal = {Adv. Mater.},
	title = {Unconventional Spin State Driven Spontaneous Magnetization in a Praseodymium Iron Antimonide},
	volume={35},
	number={8},
	pages={2207945},
	year={2023},
	doi={https://doi.org/10.1002/adma.202207945}
}

@article{nasir2010,
	author = {Nasir, Navida and Grytsiv, Andriy and Rogl, Peter and Kaczorowski, Dariusz and Effenberger, Herta Silvia},
	doi = {10.1016/j.intermet.2010.08.037},
	issn = {09669795},
	journal = {Intermetallics},
	month = {dec},
	number = {12},
	pages = {2361--2376},
	title = {{The system {N}d-{F}e-{S}b: Phase equilibria, crystal structures and physical properties}},
	volume = {18},
	year = {2010}
}

@article{imada1998metal,
  title={Metal-insulator transitions},
  author={Imada, Masatoshi and Fujimori, Atsushi and Tokura, Yoshinori},
  journal={Rev. Mod. Phys.},
  volume={70},
  number={4},
  pages={1039},
  year={1998},
  publisher={APS},
  doi={https://doi.org/10.1103/RevModPhys.70.1039}
}

@article{salamon2001physics,
  title={The physics of manganites: Structure and transport},
  author={Salamon, Myron B and Jaime, Marcelo},
  journal={Rev. Mod. Phys.},
  volume={73},
  number={3},
  pages={583},
  year={2001},
  publisher={APS},
  doi={https://doi.org/10.1103/RevModPhys.73.583}
}

@article{brookes2018beamline,
  title={The beamline {ID}32 at the {ESRF} for soft {X}-ray high energy resolution resonant inelastic {X}-ray scattering and polarisation dependent {X}-ray absorption spectroscopy},
  author={Brookes, NB and Yakhou-Harris, F and Kummer, K and Fondacaro, A and Cezar, JC and Betto, D and Velez-Fort, E and Amorese, A and Ghiringhelli, G and Braicovich, L and Barret, R},
  journal={Nucl. Instrum. Methods Phys. Res. A},
  volume={903},
  pages={175--192},
  year={2018},
  publisher={Elsevier},
  doi={https://doi.org/10.1016/j.nima.2018.07.001}
}

@article{kummer2016high,
  title={The high-field magnet endstation for {X}-ray magnetic dichroism experiments at {ESRF} soft {X}-ray beamline {ID}32},
  author={Kummer, K and Fondacaro, A and Jimenez, E and Velez-Fort, E and Amorese, A and Aspbury, M and Yakhou-Harris, F and Van Der Linden, P and Brookes, NB},
  journal={Synchrotron Radiation},
  volume={23},
  number={2},
  pages={464--473},
  year={2016},
  publisher={International Union of Crystallography},
  doi={https://doi.org/10.1107/S160057751600179X}
}

@article{senba2016upgrade,
  title={Upgrade of beamline {BL}25{SU} for soft x-ray imaging and spectroscopy of solid using nano-and micro-focused beams at {SP}ring-8},
  author={Senba, Y and Ohashi, H and Kotani, Y and Nakamura, T and Muro, T and Ohkochi, T and Tsuji, N and Kishimoto, H and Miura, T and Tanaka, M and Higashiyama, M and Takahashi, S and Ishizawa, Y and Matsushita, T and Furukawa, Y and Ohata, T and Nariyama, N and Takeshita, K and Kinoshita, T and Fujiwara, A and Takata, M and Goto, S},
  journal={AIP Conf. Proc.},
  volume={1741},
  number={1},
  pages={030044},
  year={2016},
  doi={https://doi.org/10.1063/1.4952867}
}

@article{nakamura2005development,
  title={Development of a soft X-ray magnetic circular dichroism spectrometer using a 1.9 {T} electromagnet at {BL}25{SU} of {SP}ring-8},
  author={Nakamura, T and Muro, T and Guo, FZ and Matsushita, T and Wakita, T and Hirono, 7 T and Takeuchi, Y and Kobayashi, K},
  journal={J. Electron Spectrosc. Relat. Phenom.},
  volume={144},
  pages={1035--1038},
  year={2005},
  publisher={Elsevier},
  doi={https://doi.org/10.1016/j.elspec.2005.01.146}
}

@article{weschke2018ue46,
	title={The {UE}46 {PGM}-1 beamline at {BESSY} {II}},
	author={Weschke, Eugen and Schierle, Enrico},
	journal={J. Large-Scale Res. Facilities},
	volume={4},
	pages={A127--A127},
	year={2018},
	doi={https://doi.org/10.17815/jlsrf-4-77}
}

@article{yamamoto2024field,
	title={Field-induced magnetic transitions in the highly anisotropic ferrimagnet {E}r{F}e$_5${A}l$_7$ studied by high-field x-ray magnetic dichroism},
	author={Yamamoto, Sh. and Gorbunov, D.I. and Prokhnenko, O. and Weschke, E. and Miyata, A and Diaz-Ortega, IF and Strohm, C and Duc, F and Henriques, MS and Gazizulina, A and Uhlarz, M and Mathon, O and Andreev, A V and Nojiri, H and Wosnitza, J},
	journal={Phys. Rev. B},
	volume={109},
	number={9},
	pages={094404},
	year={2024},
	publisher={APS},
	doi={https://doi.org/10.1103/PhysRevB.109.094404}
}

@misc{NdFeSbSupplemental,
  note = {See Supplemental Material at http://link.aps.org/supplemental/... for details of sample synthesis, pulsed-field magnetization, specific heat, XAS spectra, the extraction of Nd, Fe, and Sb magnetic moments, and DFT calculations, which includes Refs. \cite{pabst2022,pabst2023synthetic,skourski2011high,kotte2022orbital,yasui2015temperature,chen1995experimental,kummer2013thin,jiang2013experimental,gurevich1999nexafs,ye2015carrier,thole1992x,carra1993x,teramura1996effect,fernandez2020influence,stoll2016magnetic,cid2017perpendicular,tcakaev2023intermixing,PBE1996,Koepernik1999,lejaeghere2016reproducibility,duffy2018imposing}.}
}

@article{ye2015carrier,
  title={Carrier-mediated ferromagnetism in the magnetic topological insulator {C}r-doped ({S}b, {B}i)$_2${T}e$_3$},
  author={Ye, Mao and Li, Wei and Zhu, Siyuan and Takeda, Yukiharu and Saitoh, Yuji and Wang, Jiajia and Pan, Hong and Nurmamat, Munisa and Sumida, Kazuki and Ji, Fuhao and Liu, Z},
  journal={Nat. Commun.},
  volume={6},
  number={1},
  pages={8913},
  year={2015},
  publisher={Nature Publishing Group UK London},
  doi={https://doi.org/10.1038/ncomms9913}
}

@article{kimura1997magnetic,
  title={Magnetic circular dichroism in the soft-x-ray absorption spectra of {M}n-based magnetic intermetallic compounds},
  author={Kimura, Akio and Suga, Shigemasa and Shishidou, Tatsuya and Imada, Shin and Muro, Takayuki and Park, Serng-Yerl and Miyahara, Tsuneaki and Kaneko, Takejiro and Kanomata, Takeshi},
  journal={Phys. Rev. B},
  volume={56},
  number={10},
  pages={6021},
  year={1997},
  publisher={APS},
  doi={https://doi.org/10.1103/PhysRevB.56.6021}
}

@article{holm2002xmcd,
  title={XMCD characterization of the ferromagnetic state of {Y}b$_{14}${M}n{S}b$_{11}$},
  author={Holm, Aaron P and Kauzlarich, Susan M and Morton, Simon A and Waddill, G Dan and Pickett, Warren E and Tobin, James G},
  journal={J. Am. Chem. Soc.},
  volume={124},
  number={33},
  pages={9894--9898},
  year={2002},
  publisher={ACS Publications},
  doi={https://doi.org/10.1021/ja020564y}
}

@article{murakawa2010ferroelectricity,
  title={Ferroelectricity induced by spin-dependent metal-ligand hybridization in {B}a$_2${C}o{G}e$_2${O}$_7$},
  author={Murakawa, H and Onose, Yoshinori and Miyahara, S and Furukawa, N and Tokura, Y},
  journal={Phys. Rev. Lett.},
  volume={105},
  number={13},
  pages={137202},
  year={2010},
  publisher={APS},
  doi={https://doi.org/10.1103/PhysRevLett.105.137202}
}

@article{yamagami2021itinerant,
  title={Itinerant ferromagnetism mediated by giant spin polarization of the metallic ligand band in the van der Waals magnet {F}e$_5${G}e{T}e$_2$},
  author={Yamagami, Kohei and Fujisawa, Yuita and Driesen, Benjamin and Hsu, Chia-Hsiu and Kawaguchi, Kaisyu and Tanaka, Hiroki and Kondo, Takeshi and Zhang, Yujun and Wadati, Hiroki and Araki, Kosei and Takeda, T},
  journal={Phys. Rev. B},
  volume={103},
  number={6},
  pages={L060403},
  year={2021},
  publisher={APS},
  doi={https://doi.org/10.1103/PhysRevB.103.L060403}
}

@article{wang2024double,
	title={Double exchange interaction in {M}n-based topological kagome ferrimagnet},
	author={Wang, Jiameng and Ernst, Arthur and Antonov, Victor N and Jiang, Qi and Qian, Haoji and Wang, Deyang and Cao, Jiefeng and Zhu, Fangyuan and Qiao, Shan and Ye, Mao},
	journal={Commun. Phys.},
	volume={7},
	number={1},
	pages={350},
	year={2024},
	publisher={Nature Publishing Group UK London},
	doi={https://doi.org/10.1038/s42005-024-01838-9}
}

@article{kotte2022orbital,
  title={Orbital-induced crossover of the {F}ulde-{F}errell-{L}arkin-{O}vchinnikov phase into {A}brikosov-like states},
  author={Kotte, Tommy and K{\"u}hne, Hannes and Schlueter, John A and Zwicknagl, Gertrud and Wosnitza, J},
  journal={Phys. Rev. B},
  volume={106},
  number={6},
  pages={L060503},
  year={2022},
  publisher={APS},
  doi={https://doi.org/10.1103/PhysRevB.106.L060503}
}

@article{yasui2015temperature,
  title={Temperature dependence of post-sintered annealing on magnetic properties of intergranular phase in {N}d-{F}e-{B} permanent magnet},
  author={Yasui, Akira and Nakamura, Tetsuya and Kotani, Yoshinori and Fukagawa, Tomoki and Nishiuchi, Takeshi and Hirosawa, Satoshi},
  journal={J. Appl. Phys.},
  volume={117},
  number={17},
  pages={17B313},
  year={2015},
  publisher={AIP Publishing},
  doi={https://doi.org/10.1063/1.4916938}
}

@article{chen1995experimental,
  title={Experimental confirmation of the X-ray magnetic circular dichroism sum rules for iron and cobalt},
  author={Chen, CT and Idzerda, YU and Lin, H-J and Smith, NV and Meigs, G and Chaban, E and Ho, GH and Pellegrin, E and Sette, F},
  journal={Phys. Rev. Lett.},
  volume={75},
  number={1},
  pages={152},
  year={1995},
  publisher={APS},
  doi={https://doi.org/10.1103/PhysRevLett.75.152}
}

@article{thole1992x,
	title={X-ray circular dichroism as a probe of orbital magnetization},
	author={Thole, B.T. and Carra, P. and Sette, F. and van der Laan, G.},
	journal={Phys. Rev. Lett.},
	volume={68},
	number={12},
	pages={1943},
	year={1992},
	publisher={APS},
	doi={https://doi.org/10.1103/PhysRevLett.68.1943}
}

@article{carra1993x,
	title={{X}-ray circular dichroism and local magnetic fields},
	author={Carra, P. and Thole, B.T. and Altarelli, M. and Wang, X.},
	journal={Phys. Rev. Lett.},
	volume={70},
	number={5},
	pages={694},
	year={1993},
	publisher={APS},
	doi={https://doi.org/10.1103/PhysRevLett.70.694}
}

@article{teramura1996effect,
  title={Effect of {C}oulomb interaction on the {X}-ray magnetic circular dichroism spin sum rule in rare earths},
  author={Teramura, Y and Tanaka, A and Jo, T},
  journal={J. Phys. Soc. Jpn.},
  volume={65},
  number={9},
  pages={3056--3059},
  year={1996},
  publisher={The Physical Society of Japan},
  doi={https://doi.org/10.1143/JPSJ.65.3056}
}

@article{cid2017perpendicular,
  title={Perpendicular magnetic anisotropy in amorphous {N}d$_{x}${C}o$_{1-x}$ thin films studied by x-ray magnetic circular dichroism},
  author={Cid, R and Alameda, JM and Valvidares, SM and Cezar, JC and Bencok, P and Brookes, NB and D{\'\i}az, J},
  journal={Phys. Rev. B},
  volume={95},
  number={22},
  pages={224402},
  year={2017},
  publisher={APS},
  doi={https://doi.org/10.1103/PhysRevB.95.224402}
}

@article{nguyen2024ligand,
  title={Ligand Field Calculations for Element-Specific Spectroscopy of Rare Earth--Transition Metal Ferrimagnetic Alloys},
  author={Nguyen, Truc Ly and Mazet, Thomas and Chang, Fan-Hsiu and Lin, Hong-Ji and Chen, Chien-Te and Chainani, Ashish},
  journal={J. Phys. Soc. Jpn.},
  volume={93},
  number={12},
  pages={121006},
  year={2024},
  publisher={The Physical Society of Japan},
  doi={https://doi.org/10.7566/JPSJ.93.121006}
}

@article{thompson2014synthesis,
  title={Synthesis, structures, and magnetic properties of rare-earth cobalt arsenides, {RC}o$_2${A}s$_2$ ({R}= {L}a, {C}e, {P}r, {N}d)},
  author = {Thompson, Corey M and Tan, Xiaoyan and Kovnir, Kirill and Garlea, V Ovidiu and Gippius, Andrei A and Yaroslavtsev, Alexander A and Menushenkov, Alexey P and Chernikov, Roman V and Büttgen, Norbert and Krätschmer, Wolfgang and Zubavichus, Yan V and Shatruk, Michael},
  journal={Chem. Mater.},
  volume={26},
  number={12},
  pages={3825--3837},
  year={2014},
  publisher={ACS Publications},
  doi={https://doi.org/10.1021/cm501522v}
}

@article{ohta2016anomalous,
  title={Anomalous Behavior of Localized Magnetic Moments in Itinerant Ferromagnets {L}n$_2${C}o$_{12}${P}$_7$ ({L}n= {Y}, {P}r, {N}d, {S}m, {G}d and {D}y)},
  author={Ohta, Hiroto and Watanabe, Yusuke and Miyake, Atsushi and Tokunaga, Masashi and Katori, Hiroko Aruga},
  journal={J. Jpn. Soc. Powder Powder Metall.},
  volume={63},
  number={7},
  pages={652--656},
  year={2016},
  publisher={Japan Society of Powder and Powder Metallurgy},
  doi={https://doi.org/10.2497/jjspm.63.652}
}

@article{tan2016synthesis,
  title={Synthesis, crystal structure, and magnetism of {A}$_2${C}o$_{12}${A}s$_7$ ({A}= {C}a, {Y}, {C}e--{Y}b)},
  author={Tan, Xiaoyan and Garlea, V Ovidiu and Chai, Ping and Geondzhian, Andrey Y and Yaroslavtsev, Alexander A and Xin, Yan and Menushenkov, Alexey P and Chernikov, Roman V and Shatruk, Michael},
  journal={J. Solid State Chem.},
  volume={236},
  pages={147--158},
  year={2016},
  publisher={Elsevier},
  doi={https://doi.org/10.1016/j.jssc.2015.08.038}
}

@article{skourski2011high,
  title={High-field magnetization of {H}o$_2${F}e$_{17}$},
  author={Skourski, Y and Kuz’min, MD and Skokov, KP and Andreev, AV and Wosnitza, J},
  journal={Phys. Rev. B},
  volume={83},
  number={21},
  pages={214420},
  year={2011},
  publisher={APS},
  doi={https://doi.org/10.1103/PhysRevB.83.214420}
}

@phdthesis{pabst2023synthetic,
  title        = {{\normalfont Synthetic\ Challenges\ in\ Rare\ Earth\ Iron\ Antimonides\ and\ Bismuth\ Bromide\ Systems}},
  author       = {Pabst, Falk},
  year         = {2023},
  school       = {Technische Universit{\"a}t Dresden},
  type         = {{P}h.{D.} thesis},
  url={https://nbn-resolving.org/urn:nbn:de:bsz:14-qucosa2-851770}
}

@article{Koepernik1999,
  title = {Full-potential nonorthogonal local-orbital minimum-basis band-structure scheme},
  author = {Koepernik, Klaus and Eschrig, Helmut},
  journal = {Phys. Rev. B},
  volume = {59},
  issue = {3},
  pages = {1743--1757},
  numpages = {0},
  year = {1999},
  month = {Jan},
  publisher = {American Physical Society},
  doi = {10.1103/PhysRevB.59.1743},
  url = {https://link.aps.org/doi/10.1103/PhysRevB.59.1743}
}

@article{PBE1996,
  title = {Generalized Gradient Approximation Made Simple},
  author = {Perdew, John P. and Burke, Kieron and Ernzerhof, Matthias},
  journal = {Phys. Rev. Lett.},
  volume = {77},
  issue = {18},
  pages = {3865--3868},
  numpages = {0},
  year = {1996},
  month = {Oct},
  publisher = {American Physical Society},
  doi = {10.1103/PhysRevLett.77.3865},
  url = {https://link.aps.org/doi/10.1103/PhysRevLett.77.3865}
}

@article{jiang2013experimental,
  title={Experimental and theoretical investigation of the electronic structure of {C}u$_2${O} and {C}u{O} thin films on {C}u(110) using x-ray photoelectron and absorption spectroscopy},
  author={Jiang, Peng and Prendergast, David and Borondics, Ferenc and Porsgaard, Soeren and Giovanetti, Lisandro and Pach, Elzbieta and Newberg, John and Bluhm, Hendrik and Besenbacher, Flemming and Salmeron, Miquel},
  journal={J. Chem. Phys.},
  volume={138},
  number={2},
  pages={024704},
  year={2013},
  publisher={AIP Publishing},
  url = {https://doi.org/10.1063/1.4773583}
}

@article{gurevich1999nexafs,
  title={A {NEXAFS} investigation of the formation and decomposition of {C}u{O} and {C}u$_2${O} thin films on {C}u(100)},
  author={Gurevich, Alejandra B and Bent, Brian E and Teplyakov, Andrew V and Chen, Jingguang G},
  journal={Surf. Sci.},
  volume={442},
  number={1},
  pages={L971--L976},
  year={1999},
  publisher={Elsevier},
  doi={https://doi.org/10.1016/S0039-6028(99)00913-9}
}

@article{fernandez2020influence,
  title={Influence of 4f filling on electronic and magnetic properties of rare earth-Au surface compounds},
  author={Fernandez, Laura and Blanco-Rey, Mar{\'\i}a and Castrillo-Bodero, Rodrigo and Ilyn, Maxim and Ali, Khadiza and Turco, Elia and Corso, Martina and Ormaza, Maider and Gargiani, Pierluigi and Valbuena, Miguel A and others},
  journal={Nanoscale},
  volume={12},
  number={43},
  pages={22258--22267},
  year={2020},
  publisher={Royal Society of Chemistry},
  doi={10.1039/D0NR04964F}
}

@article{stoll2016magnetic,
  title={Magnetic anisotropy in surface-supported single-ion lanthanide complexes},
  author={Stoll, Paul and Bernien, Matthias and Rolf, Daniela and Nickel, Fabian and Xu, Qingyu and Hartmann, Claudia and Umbach, Tobias R and Kopprasch, Jens and Ladenthin, Janina N and Schierle, Enrico and others},
  journal={Phys. Rev. B},
  volume={94},
  number={22},
  pages={224426},
  year={2016},
  publisher={APS},
  doi={https://doi.org/10.1103/PhysRevB.94.224426}
}

@article{jackson2006effect,
  title={Effect of rare-earth doping in {$R$}{C}r{S}b$_3$ ({$R$}= {L}a, {P}r, {S}m, and {G}d)},
  author={Jackson, DD and Fisk, Z},
  journal={Phys. Rev. B},
  volume={73},
  number={2},
  pages={024421},
  year={2006},
  publisher={APS},
  doi={https://doi.org/10.1103/PhysRevB.73.024421}
}

@article{inamdar2008anisotropic,
  title={Anisotropic magnetism in {P}r{C}r{S}b3: {B}ulk properties of single crystals},
  author={Inamdar, Manjusha and Thamizhavel, A and Ramakrishnan, S},
  journal={Phys. Rev. B},
  volume={77},
  number={13},
  pages={132410},
  year={2008},
  publisher={APS},
  doi={https://doi.org/10.1103/PhysRevB.77.132410}
}

@inproceedings{inamdar2009interplay,
  title={Interplay of 3d and 4f magnetism in {RC}r{S}b$_3$},
  author={Inamdar, Manjusha and Ramakrishnan, S and Thamizhavel, A and Nigam, AK},
  booktitle={J. Phys.: Conf. Ser.},
  volume={150},
  number={4},
  pages={042163},
  year={2009},
  doi={10.1088/1742-6596/150/4/042163}
}

@article{inamdar2009anisotropic,
  title={Anisotropic magnetism in {P}r{C}r{S}b$_3$ and {N}d{C}r{S}b$_3$},
  author={Inamdar, Manjusha and Thamizhavel, A and Ramakrishnan, S},
  journal={J. Alloys Compd.},
  volume={480},
  number={1},
  pages={28--30},
  year={2009},
  publisher={Elsevier},
  doi={https://doi.org/10.1016/j.jallcom.2008.09.186}
}

@article{toulemonde1998orientation,
  title={Orientation of Nd and Mn magnetic moments in a {CMR} {N}d$_{0.72}${B}a$_{0.28}${M}n{O}$_3$ by {X}-ray magnetic circular dichroism},
  author={Toulemonde, O and Studer, F and Llobet, A and Ranno, L and Maignan, A and Pollert, E and Nevriva, M and Pellegrin, E and Brooks, NB and Goedkoop, J},
  journal={J. Magn. Magn. Mater.},
  volume={190},
  number={3},
  pages={307--317},
  year={1998},
  publisher={Elsevier},
  doi={https://doi.org/10.1016/S0304-8853(98)00296-0}
}

@article{munoz2000magnetic,
  title={Magnetic structure evolution of {N}d{M}n{O}$_3$ derived from neutron diffraction data},
  author={Munoz, A and Alonso, JA and Martinez-Lope, MJ and Garc{\'\i}a-Mu{\~n}oz, JL and Fern{\'a}ndez-D{\'\i}az, MT},
  journal={J. Phys.: Condens. Matter},
  volume={12},
  number={7},
  pages={1361--1376},
  year={2000},
  doi={10.1088/0953-8984/12/7/319}
}

@article{kumar2025exchange,
  title={Exchange interactions, anisotropy, and low-temperature magnetization reversal in {N}d{F}e{O}$_3$},
  author={Kumar, Amit and Bera, AK and Balz, Christian and da Silva, Ivan and Yusuf, SM},
  journal={Phys. Rev. B},
  volume={112},
  number={9},
  pages={094444},
  year={2025},
  publisher={APS},
  doi={https://doi.org/10.1103/8qy2-k5cb}
}

@article{pal2019peculiar,
  title={Peculiar magnetic states in the double perovskite {N}d$_2${N}i{M}n{O}$_6$},
  author={Pal, Somnath and Jana, Somnath and Govinda, Sharada and Pal, Banabir and Mukherjee, Sumanta and Keshavarz, Samara and Thonig, Danny and Kvashnin, Yaroslav and Pereiro, Manuel and Mathieu, Roland and others},
  journal={Phys. Rev. B},
  volume={100},
  number={4},
  pages={045122},
  year={2019},
  publisher={APS},
  doi={https://doi.org/10.1103/PhysRevB.100.045122}
}

@article{lejaeghere2016reproducibility,
  title={Reproducibility in density functional theory calculations of solids},
  author={Lejaeghere, Kurt and Bihlmayer, Gustav and Bj{\"o}rkman, Torbj{\"o}rn and Blaha, Peter and Bl{\"u}gel, Stefan and Blum, Volker and Caliste, Damien and Castelli, Ivano E and Clark, Stewart J and Dal Corso, Andrea and others},
  journal={Science},
  volume={351},
  number={6280},
  pages={aad3000},
  year={2016},
  publisher={American association for the advancement of science},
  doi={10.1126/science.aad3000}
}

@article{duffy2018imposing,
  title={Imposing long-range ferromagnetic order in rare-earth-doped magnetic topological-insulator heterostructures},
  author={Duffy, LB and Frisk, A and Burn, DM and Steinke, N-J and Herrero-Martin, J and Ernst, Arthur and Van Der Laan, G and Hesjedal, T},
  journal={Phys. Rev. Mater.},
  volume={2},
  number={5},
  pages={054201},
  year={2018},
  publisher={APS},
  doi={https://doi.org/10.1103/PhysRevMaterials.2.054201}
}

@article{boada2010decoupling,
  title={Decoupling of the magnetic sublattices at the compensation point in $R$-{F}e compounds},
  author={Boada, R and Piquer, Cristina and Laguna-Marco, MA and Chaboy, Jes{\'u}s},
  journal={Phys. Rev. B},
  volume={82},
  number={5},
  pages={052407},
  year={2010},
  publisher={APS},
  doi={https://doi.org/10.1103/PhysRevB.82.052407}
}

@incollection{franse1993magnetic,
  author    = {Franse, J. J. M. and Radwa\'{n}ski, R. J.},
  title     = {Magnetic properties of binary rare-earth 3d-transition-metal intermetallic compounds},
  booktitle = {Handbook of Magnetic Materials},
  editor    = {Buschow, K. H. J.},
  volume    = {7},
  publisher = {Elsevier},
  address   = {Amsterdam},
  year      = {1993},
  pages     = {307}
}

@article{tcakaev2023intermixing,
  title={Intermixing-{D}riven {S}urface and {B}ulk {F}erromagnetism in the {Q}uantum {A}nomalous {H}all {C}andidate {M}n{B}i$_6${T}e$_{10}$},
  author={Tcakaev, Abdul-Vakhab and Rubrecht, Bastian and Facio, Jorge I and Zabolotnyy, Volodymyr B and Corredor, Laura T and Folkers, Laura C and Kochetkova, Ekaterina and Peixoto, Thiago RF and Kagerer, Philipp and Heinze, Simon and others},
  journal={Adv. Sci.},
  volume={10},
  number={10},
  pages={2203239},
  year={2023},
  publisher={Wiley Online Library},
  doi={https://doi.org/10.1002/advs.202203239}
}

@article{kummer2013thin,
  title={Thin conductive diamond films as beam intensity monitors for soft x-ray beamlines},
  author={Kummer, K and Fondacaro, A and Yakhou-Harris, F and Sessi, V and Pobedinskas, Paulius and Janssens, SD and Haenen, Ken and Williams, Oliver Aneurin and Hees, J and Brookes, NB},
  journal={Rev. Sci. Instrum.},
  volume={84},
  number={3},
  pages={035105},
  year={2013},
  publisher={AIP Publishing},
  doi={https://doi.org/10.1063/1.4794439}
}

\end{document}


\title{Supplemental Material to ``Sb-5$p$-mediated antiparallel Nd--Fe coupling in Nd$_3$Fe$_3$Sb$_7$''}

\author{S. Palazzese}
\affiliation{Dresden High Magnetic Field Laboratory (HLD-EMFL) and Würzburg-Dresden Cluster of Excellence ctd.qmat, Helmholtz-Zentrum Dresden-Rossendorf, 01328 Dresden, Germany}%
\affiliation{Institute of Solid State and Materials Physics, Technische Universität Dresden, 01062 Dresden, Germany}
\author{Sh. Yamamoto}
\email{s.yamamoto@hzdr.de}
\affiliation{Dresden High Magnetic Field Laboratory (HLD-EMFL) and Würzburg-Dresden Cluster of Excellence ctd.qmat, Helmholtz-Zentrum Dresden-Rossendorf, 01328 Dresden, Germany}%
\author{T. Kotte}
\affiliation{Dresden High Magnetic Field Laboratory (HLD-EMFL) and Würzburg-Dresden Cluster of Excellence ctd.qmat, Helmholtz-Zentrum Dresden-Rossendorf, 01328 Dresden, Germany}%
\author{R. Ray}
\affiliation{Leibniz IFW Dresden, Helmholtzstr. 20, 01069 Dresden, Germany}
\affiliation{Department of Physics, Birla Institute of Technology Mesra, Ranchi 835215, India}%
\author{F. Pabst}
\affiliation{Faculty of Chemistry and Food Chemistry and Würzburg-Dresden Cluster of Excellence ctd.qmat, Technische Universität Dresden, 01062 Dresden, Germany}
\author{S. Chattopadhyay}
\affiliation{Dresden High Magnetic Field Laboratory (HLD-EMFL) and Würzburg-Dresden Cluster of Excellence ctd.qmat, Helmholtz-Zentrum Dresden-Rossendorf, 01328 Dresden, Germany}
\affiliation{UGC-DAE Consortium for Scientific Research Mumbai Centre, 246-C CFB, BARC Campus, Mumbai 400085, India}
\author{D. I. Gorbunov}
\affiliation{Dresden High Magnetic Field Laboratory (HLD-EMFL) and Würzburg-Dresden Cluster of Excellence ctd.qmat, Helmholtz-Zentrum Dresden-Rossendorf, 01328 Dresden, Germany}%
\author{T. Herrmannsdörfer}
\affiliation{Dresden High Magnetic Field Laboratory (HLD-EMFL) and Würzburg-Dresden Cluster of Excellence ctd.qmat, Helmholtz-Zentrum Dresden-Rossendorf, 01328 Dresden, Germany}%
\author{K. Kummer}
\affiliation{European Synchrotron Radiation Facility, 38043 Grenoble Cedex 9, France}%
\author{A. Kurtanidze}
\affiliation{Dresden High Magnetic Field Laboratory (HLD-EMFL) and Würzburg-Dresden Cluster of Excellence ctd.qmat, Helmholtz-Zentrum Dresden-Rossendorf, 01328 Dresden, Germany}
\affiliation{Institute of Solid State and Materials Physics, Technische Universität Dresden, 01062 Dresden, Germany}
\author{E. Weschke}
\affiliation{Helmholtz-Zentrum Berlin für Materialien und Energie, BESSY II, 12489 Berlin, Germany
}%
\author{O. Prokhnenko}
\affiliation{Helmholtz-Zentrum Berlin für Materialien und Energie, BESSY II, 12489 Berlin, Germany
}%
\author{K. Yamamoto}
\affiliation{Institute for Molecular Science, Okazaki, Aichi 444-8585, Japan}
\author{Y. Kotani}
\affiliation{Japan Synchrotron Radiation Research Institute, SPring-8, Sayo, Hyogo 679-5198, Japan}
\author{H. Nojiri}
\affiliation{Institute for Materials Research, Tohoku University, Sendai, 980-8577, Japan}%
\author{M. Ruck}
\affiliation{Faculty of Chemistry and Food Chemistry and Würzburg-Dresden Cluster of Excellence ctd.qmat, Technische Universität Dresden, 01062 Dresden, Germany}
\author{Manuel Richter}
\affiliation{Leibniz IFW Dresden, Helmholtzstr. 20, 01069 Dresden, Germany}
\affiliation{Dresden Center for Computational Materials Science (DCMS), TU Dresden, 01062 Dresden, Germany}
\author{J. Wosnitza}
\affiliation{Dresden High Magnetic Field Laboratory (HLD-EMFL) and Würzburg-Dresden Cluster of Excellence ctd.qmat, Helmholtz-Zentrum Dresden-Rossendorf, 01328 Dresden, Germany}%
\affiliation{Institute of Solid State and Materials Physics, Technische Universität Dresden, 01062 Dresden, Germany}%

\date{\today}
\maketitle

\section{Sample preparation and characterization}

Single crystals of Nd$_3$Fe$_3$Sb$_7$ were synthesized by Bi-flux growth following exactly the protocol established for Pr$_3$Fe$_3$Sb$_7$ (identical charge/flux handling in an Ar glove box, the same time–temperature profile, hot centrifugation, and wet-chemical flux removal) as detailed in Refs. \cite{pabst2022,pabst2023synthetic}.

\begin{figure}[tb]
	\begin{center}
		\includegraphics[width=0.85\linewidth]{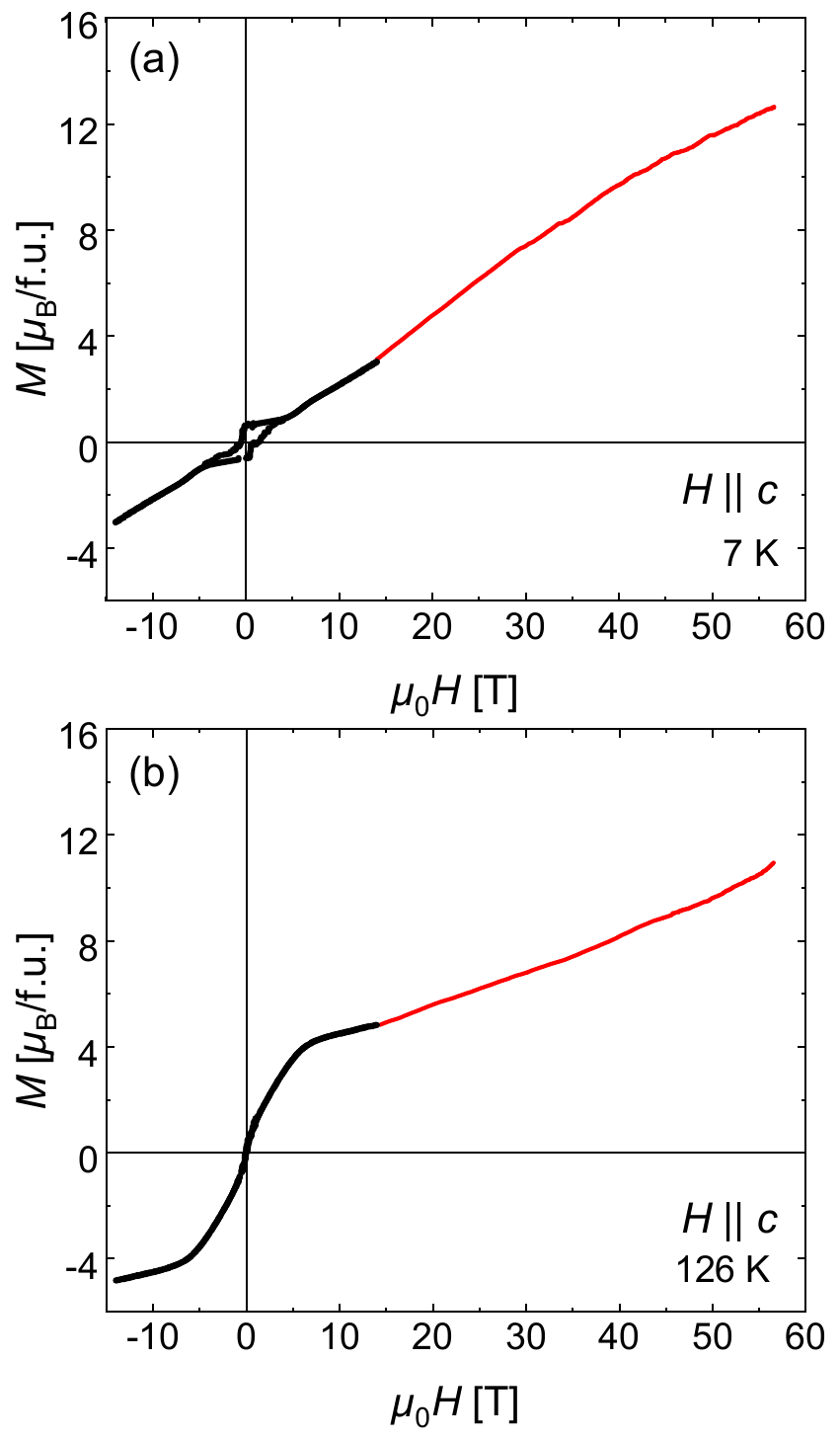}
	\end{center}
	\caption{Field-dependent magnetization of Nd$_3$Fe$_3$Sb$_7$ with magnetic field applied along the $c$ axis at (a) 7 K and (b) 126 K. Pulsed-field data are shown in red, and static-field measurements are shown in black.}
	\label{Fig:HighMH}
\end{figure}

Phase purity and structure were checked by powder x-ray diffraction; Nd$_3$Fe$_3$Sb$_7$ is isostructural to Pr$_3$Fe$_3$Sb$_7$ (hexagonal $P6_3/m$). Elemental analysis by energy-dispersive x-ray spectroscopy on polished crystals shows a homogeneous distribution of Nd, Fe, and Sb \cite{pabst2023synthetic}. 

\section{Static-field magnetization and magnetic anisotropy}

For the magnetization measurements, particular care was taken to control the magnetic-field history in the lowest-field measurements. Here, zero field denotes the residual-field condition, $|\mu_0 H| \leq 10^{-4}$ T. For the zero-field and 0.01 T SQUID measurements shown in Figs. 1(c) and 1(d), the magnet had been kept off in the warm state before the 0 T measurement. The sample had already been centered in previous measurements, and no magnetic field was applied for recentering immediately before the 0 T measurement. The temperature-dependent magnetization was first measured at 0 T, and the field was subsequently increased to 0.01 T for the low-field measurement. After this sequence, the sample centering was checked again and found to be maintained.

The temperature- and field-dependent SQUID magnetization data shown in Figs. 1(c)--1(f) of the main text demonstrate the strong magnetic anisotropy in Nd$_3$Fe$_3$Sb$_7$. Above $T_\mathrm{sr}$, the low-field magnetization is much larger for $H \perp c$ than for $H \parallel c$, evidencing a larger magnetic susceptibility within the $ab$ plane in this temperature range. Below $T_\mathrm{sr}$, a finite remanent magnetization also develops for $H \parallel c$, consistent with the emergence of a $c$-axis component through the spin reorientation. The inset of Fig. 1(d) shows that $T_\mathrm{sr}$ depends on both the magnitude and direction of the applied field, as expected for a transition governed by magnetic anisotropy. The present measurements do not establish $T_\mathrm{sr}$ as the onset of magnetic ordering of the Nd sublattice. In particular, the field-dependent Nd XMCD response observed at 100 K [Figs. 2(c) and 3(b) of the main text] does not allow us to distinguish whether the Nd sublattice is already magnetically ordered above $T_\mathrm{sr}$ or whether the Nd moments are polarized by the exchange field of the Fe sublattice.

The negative low-field magnetization observed for $H \parallel c$ should not be interpreted simply as an equilibrium magnetization following the applied field through the compensation point. In a ferrimagnet, the compensation temperature is generally observed as a minimum in $M(T)$; the negative value at low fields arises when the magnetic anisotropy and coercivity hinder a reversal of the net magnetization into the field direction, leaving the magnetization blocked opposite to the applied field.

At 1.8 K, the finite remanent magnetizations of 0.48 $\mu_\mathrm{B}$/f.u. for $H \parallel c$ and 0.13 $\mu_\mathrm{B}$/f.u. for $H \perp c$ [Figs. 1(e) and 1(f) of the main text] are consistent with a low-temperature ferrimagnetic state, in which the net magnetization is not confined strictly to either the $c$ axis or the $ab$ plane. This behavior resembles that reported for the isostructural compound Pr$_3$Fe$_3$Sb$_7$, where we identified a strong magnetic anisotropy, a spin-reorientation transition, and a compensation point by magnetometry studies \cite{pabst2022}.

\section{High-field magnetization}

To investigate the field dependence of the magnetic state in Nd$_3$Fe$_3$Sb$_7$, we performed pulsed-field magnetization measurements up to 58 T applied along the $c$ axis using a coaxial pickup-coil system \cite{skourski2011high}. Figure \ref{Fig:HighMH} displays magnetization data taken at 7 and 126 K, combining data from static-field (black) and pulsed-field (red) measurements. The static-field data were obtained using vibrating-sample magnetometer (VSM).

At both temperatures, the magnetization increases monotonously with field without signs of saturation up to the highest applied field. This behavior is consistent with strong antiferromagnetic coupling between the Fe and Nd sublattices and the presence of noncollinear  magnetic configurations inferred from XMCD and magnetization data. These results indicate the robustness of the antiferromagnetic Nd--Fe intersublattice coupling and the significant anisotropy in the system, as a fully saturated collinear state is not reached even under very large magnetic fields applied along the $c$ axis.

The magnetization of 11–12 $\mu_\mathrm{B}$/f.u. reached near 50 T (Fig. \ref{Fig:HighMH}) is reasonable in view of the available moment scale. The free-ion saturation moment of Nd$^{3+}$ is $g_J J^z$ = 3.27 $\mu_\mathrm{B}$ per Nd ion, corresponding to 9.8 $\mu_\mathrm{B}$/f.u. for the three Nd ions. In addition, the ordered Fe moment in the isostructural compound Pr$_3$Fe$_3$Sb$_7$ was estimated to be about 1.5–1.7 $\mu_\mathrm{B}$ per Fe from neutron diffraction and M\"{o}ssbauer spectroscopy \cite{pabst2022}, corresponding to 4.5–5.1 $\mu_\mathrm{B}$/f.u. for the three Fe ions. Therefore, a magnetization of 11–12 $\mu_\mathrm{B}$/f.u. at 50 T is still below full saturation.

The absolute magnetization values in Fig. \ref{Fig:HighMH} should be compared with the SQUID data in Fig. 1(c) with caution. The SQUID measurements in Fig. 1(c) were performed on a single well-oriented crystal, whereas the VSM and pulsed-field measurements in Fig. \ref{Fig:HighMH} were performed on several small single-crystalline pieces mounted together in order to obtain sufficient signal. Since Nd$_3$Fe$_3$Sb$_7$ exhibits a strong magnetic anisotropy, small misalignments among these pieces can affect the absolute magnetization values. We, therefore, use the data in Fig. \ref{Fig:HighMH} only to demonstrate the absence of saturation up to 58 T, while the quantitative comparison with XMCD and DFT is based on the SQUID data from the single well-oriented crystal shown in Fig. 1(c).

\begin{figure}[tb]
	\begin{center}
		\includegraphics[width=\linewidth]{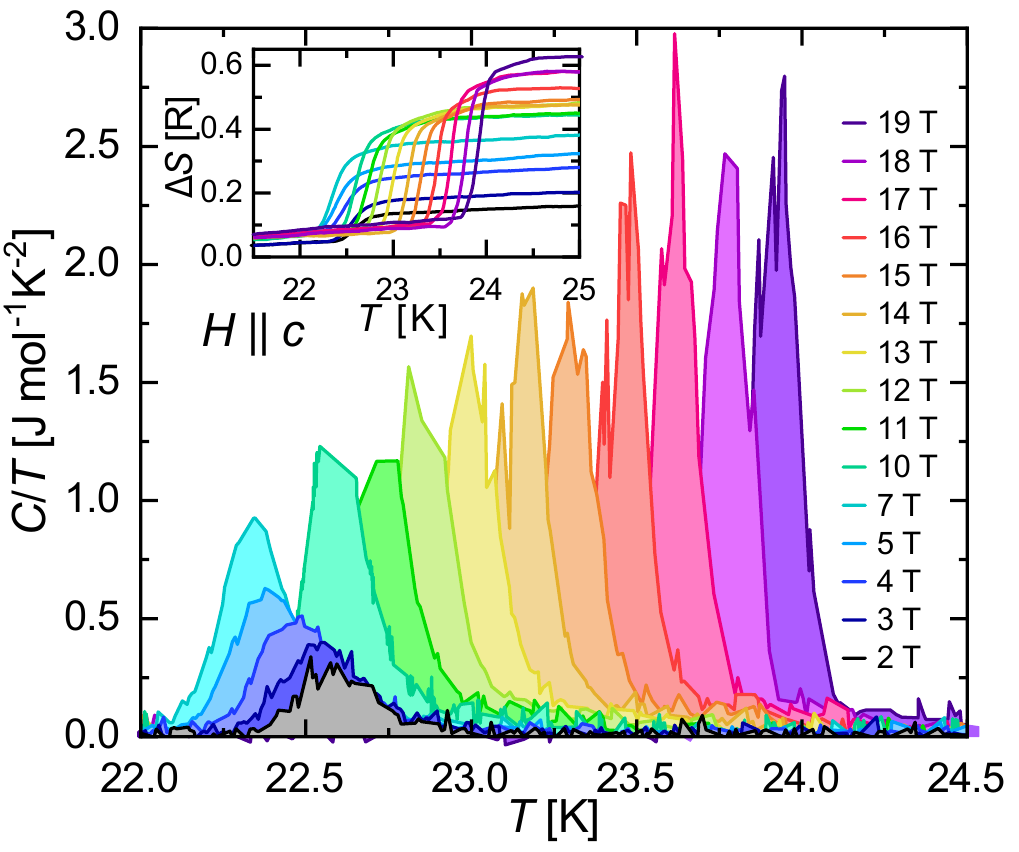}
	\end{center}
	\caption{Specific heat from continuous relaxation calorimetry in static fields. Inset: entropy change $\Delta S$ as a function of temperature. 
}
	\label{Fig:HC-SM}
\end{figure}

\section{Specific-heat results}

We performed specific-heat measurements using the continuous-relaxation method \cite{kotte2022orbital} near $T_\mathrm{comp}$. Figure \ref{Fig:HC-SM} shows that the transition temperature decreases with field up to 7 T and then increases at higher fields. The associated entropy change $\Delta S$, obtained by integrating $C/T$ in the temperature range of the measurement, rises increasingly abrupt toward higher fields (inset).  Notably, the entropy released at the transitions is less than 0.5$R$, indicating that only a small fraction of the total magnetic entropy participates in this transition. This is consistent with the narrow measurement range, which does not capture other contributions, such as those from high-temperature crystal-field excitations or low-temperature spin fluctuations.

\section{X-ray absorption spectra}

\begin{figure}[tb]
	\begin{center}
		\includegraphics[width=\linewidth]{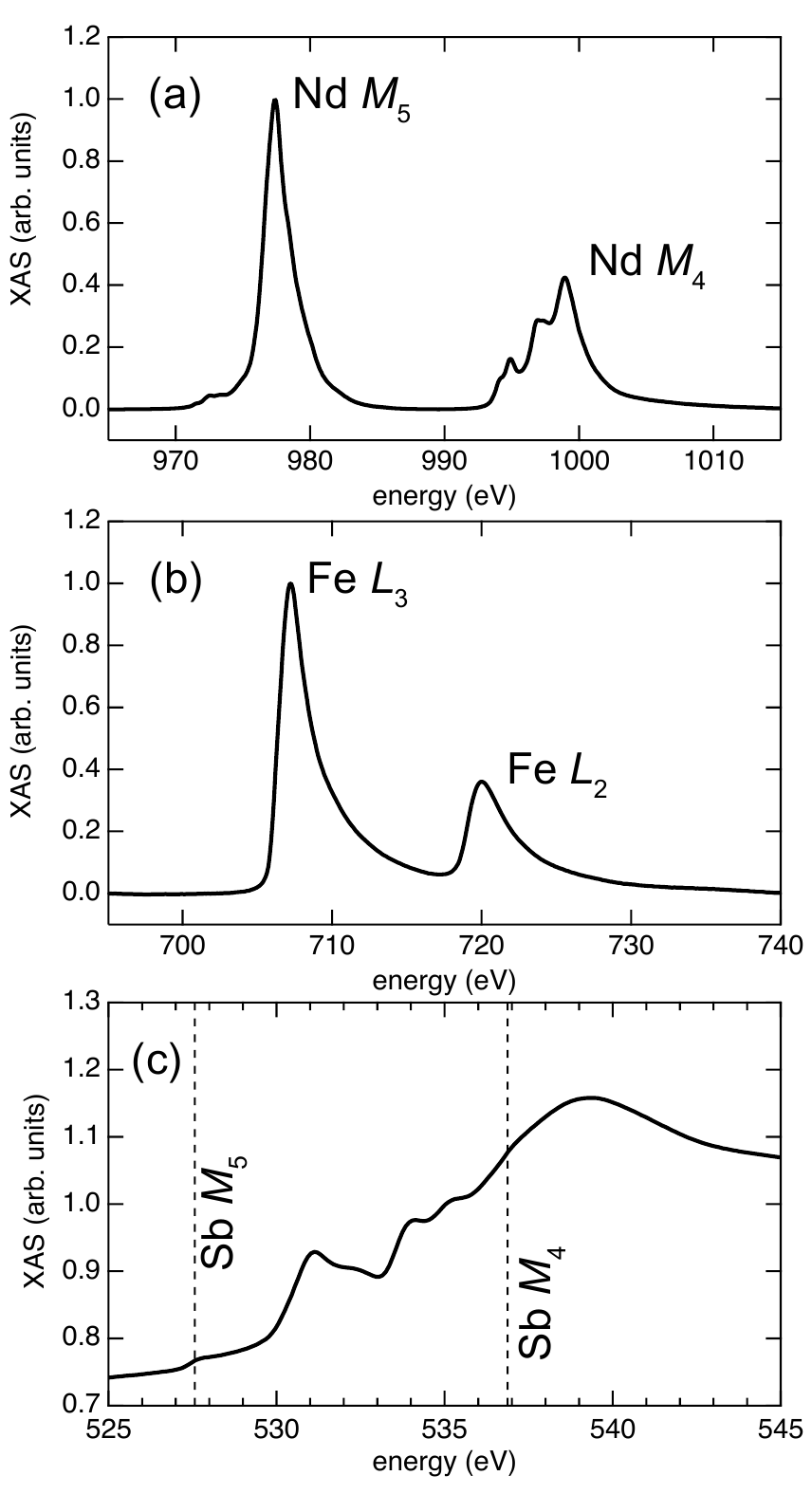}
	\end{center}
	\caption{X-ray absorption spectra measured at 100 K in a field of 9 T applied along the $c$ axis at the (a) Nd-$M_{4,5}$, (b) Fe-$L_{2,3}$, and (c) Sb-$M_{4,5}$ edges.}
	\label{Fig:XAS}
\end{figure}

Figure \ref{Fig:XAS} shows the averaged XAS spectra [($\mu^{+}$ + $\mu^{-}$)/2] for Nd, Fe, and Sb, measured at 100 K in a magnetic field of 9 T applied along the $c$ axis. Here, $\mu^{+}$ and $\mu^{-}$ denote the absorption spectra for right- and left-circularly polarized light, respectively. For the XMCD measurements at BESSY-II and ESRF, the samples were cleaved $in$ $situ$ parallel to the $ab$ plane at 10 K under UHV conditions better than 1$\times$10$^{-9}$ mbar in order to obtain clean surfaces. The line shape of the Nd $M$-edge XAS agrees well with that observed for Nd$^{3+}$ \cite{yasui2015temperature}. The line shape at the Fe $L$-edge is characteristic of metallic iron \cite{chen1995experimental}. For the ESRF measurements, the relevant lateral dimension of the cleaved surface was approximately 180 $\mu$m, while the central x-ray beam had a vertical Gaussian FWHM of approximately 30 $\mu$m. Most of the direct beam was, therefore, incident on the sample. The incident photon flux was normalized using an Au mesh. Diffraction by the mesh produces weak fringes outside the central beam \cite{kummer2013thin}. For our measurement conditions, these fringes extended approximately ±200 $\mu$m in the vertical direction. Part of the fringe intensity, therefore, illuminated the Cu sample holder outside the sample area. The lineshapes of the Fe $L_3$-edge XAS spectra measured immediately after cleavage at 10 K and 14.5 hours later at 100 K were essentially unchanged, with no indication of surface oxidation during this period.

In the Sb XAS spectra, large structures from the O $K$-edge are superimposed on the Sb $M_{4,5}$-edge region. Based on the beam-footprint considerations above and the absence of an appreciable change in the Fe XAS after 14.5 hours, we attribute these structures to the naturally oxidized Cu sample holder, which was partially illuminated by the Au-mesh diffraction fringes, rather than to oxidation of the Nd$_3$Fe$_3$Sb$_7$ surface. Reference O $K$-edge XAS spectra of copper oxides show several features in the 530--545 eV range, arising from O 2$p$ states hybridized with Cu 3$d$ and Cu 4$sp$ states \cite{jiang2013experimental,gurevich1999nexafs}. The structures around 531--535 eV and the broad feature around 539 eV in Fig. \ref{Fig:XAS}(c) can be consistently assigned to the O $K$-edge fine structure of Cu-oxide species, most likely mixed CuO/Cu$_2$O, on the oxidized Cu sample holder, rather than to intrinsic Sb absorption. The XMCD peaks shown in Fig. 2 in the main text coincide with the Sb $M_{4,5}$-edge energy region, and the separation between the two opposite-sign XMCD peaks matches the spin-orbit splitting of the Sb 3$d$ shell, consistent with previous Sb $M_{4,5}$-edge XMCD spectra \cite{ye2015carrier,duffy2018imposing}. These observations indicate that the observed XMCD signal originates from Sb.

Sb $M_{4,5}$-edge XMCD signals are generally very weak (about 0.1\% of the XAS intensity) \cite{ye2015carrier}. Under the present measurement conditions, small residual structures can appear in the XMCD difference spectrum due to statistical noise and imperfect cancellation of the large nonmagnetic background. We, therefore, do not attribute the smaller structures that do not reverse with temperature to intrinsic Sb 5$p$ spin polarization. The discussion in the main text is based on the two main near-edge XMCD peaks at the Sb $M_5$ and $M_4$ edges, rather than on these weaker residual structures.

\section{Experimental determination of N\NoCaseChange{d}, F\NoCaseChange{e}, and S\NoCaseChange{b} magnetic moments}

We applied the magneto-optical sum rules \cite{thole1992x,carra1993x} to the Fe and Nd XAS/XMCD spectra at 10 and 100 K to extract the corresponding magnetic moments. We determined the Nd 4$f$ spin moment by taking into account the ratio $\left\langle T_z \right\rangle / \left\langle S_z \right\rangle = 0.315$, which corresponds to the ground-state expectation value of the magnetic dipole operator relative to the spin operator for a free Nd$^{3+}$ ion \cite{teramura1996effect}. This value was not treated as an adjustable parameter, but was taken from the free-ion calculation for Nd$^{3+}$ \cite{teramura1996effect,fernandez2020influence,stoll2016magnetic}. Because this correction is based on the free-ion 4$f$ configuration, the absolute Nd moment may still contain systematic uncertainty associated with crystal-field effects and magnetic anisotropy. The sign and relative field dependence of the Nd XMCD signal, however, do not depend on the precise value of this correction. We also applied a correction factor for $jj$ mixing between the 3$d_{5/2}$ and 3$d_{3/2}$ multiplets, arising from 3$d$–4$f$ interactions \cite{teramura1996effect,cid2017perpendicular}. 

For Fe at the $L_{2,3}$ edges, the spin sum rule yields the effective spin moment $m^{\mathrm{eff}}_s$ = $m_s-7\left\langle T_z \right\rangle$, where $\left\langle T_z \right\rangle$ is the expectation value of the magnetic dipole operator. Because Fe resides at a non-cubic $6h$ site in $P6_3/m$, we do not attempt to separate $\langle T_z\rangle$ and, thus, report Fe as $m_s^{\mathrm{eff}}$. Accordingly, the Fe total moment in the main text (Table I) is the effective total moment $m^{\mathrm{eff}}_{tot}$ = $m^{\mathrm{eff}}_s$ + $m_l$. For the Fe $L_{2,3}$-edge sum-rule analysis, we used a 3$d$ hole number $n_h$ = 3.31, obtained from our DFT calculation for Nd$_3$Fe$_3$Sb$_7$.

The resulting individual moments are summarized in Table I in the main text. We obtained the field dependence of the Nd and Fe magnetic moments shown in Fig. 3 by normalizing the field-dependent XMCD amplitudes using the moments determined at 9 T, as described above. 

Although the sign of the Sb $M_{4,5}$-edge XMCD provides direct information on the direction of the Sb 5$p$ spin polarization, we do not extract an absolute Sb 5$p$ magnetic moment from the XMCD sum rules. A reliable quantitative analysis would require an accurate integration of the Sb $M_{4,5}$-edge XAS intensity. In the present data, however, the strong O $K$-edge fine structure from the oxidized Cu sample holder is superimposed on the Sb $M_{4,5}$-edge XAS spectrum, as shown in Fig. \ref{Fig:XAS}(c), making such an integration unreliable. We, therefore, use the Sb XMCD data to determine the sign and field dependence of the Sb 5$p$ polarization.

\begin{figure}[tb]
	\begin{center}
		\includegraphics[width=\linewidth]{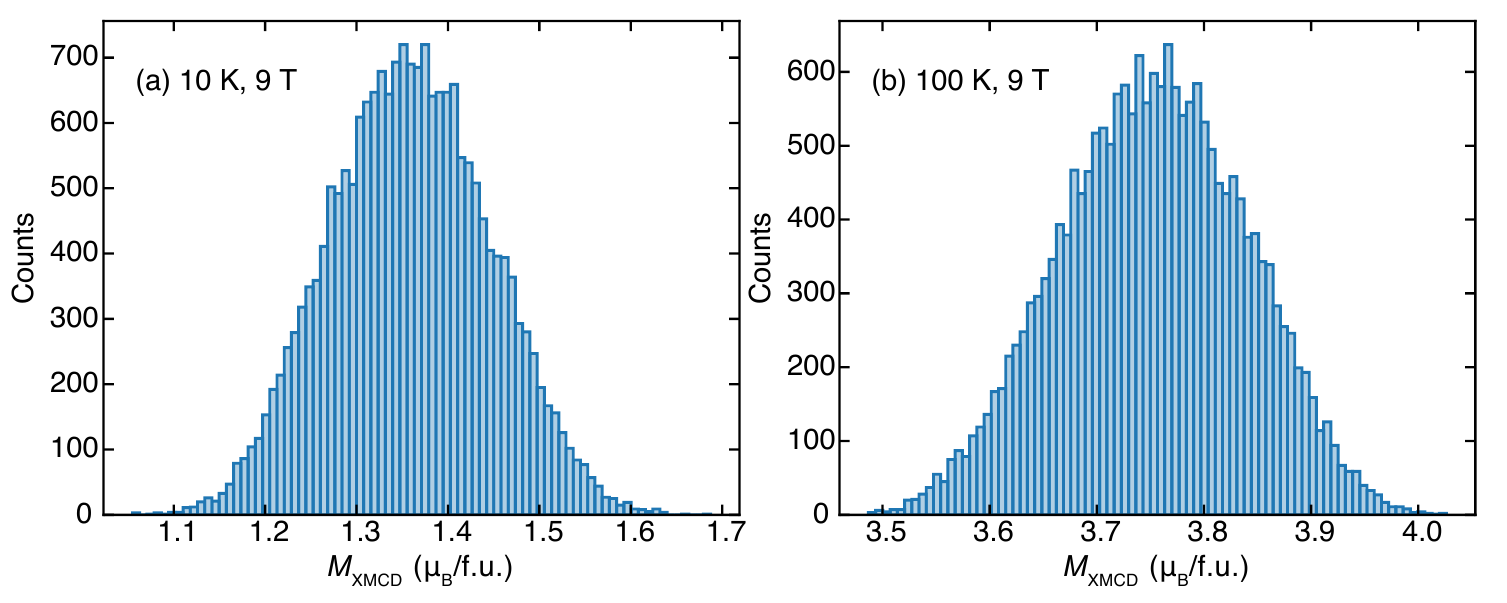}
	\end{center}
	\caption{Distributions of $M_\mathrm{XMCD}$ at (a) 10 K and (b) 100 K in a magnetic field of 9 T obtained by repeating the complete XMCD sum-rule analysis 20,000 times while randomly varying the relevant analysis parameters around the values used in the nominal analysis and including the effect of point-to-point experimental noise.}
	\label{Fig:Uncertainty}
\end{figure}

To compare the XMCD-derived moments with the bulk magnetization, we estimated the magnetization per formula unit from the XMCD sum-rule values as $M_\mathrm{XMCD}$ = $3m^{\mathrm{Nd}\text{-}4f}_{\mathrm{tot}}$ + $3m^{\mathrm{eff, Fe}\text{-}3d}_{\mathrm{tot}}$, where $m^{\mathrm{Nd}\text{-}4f}_{\mathrm{tot}}$ = $m_s^{{\mathrm{Nd}\text{-}4}f}$+$m_l^{{\mathrm{Nd}\text{-}4}f}$ and $m^{\mathrm{eff, Fe}\text{-}3d}_{\mathrm{tot}} = m_s^{{\mathrm{eff,Fe}\text{-}3}d}+m_l^{{\mathrm{Fe}\text{-}3}d}$. To evaluate the uncertainty associated with the sum-rule analysis, following the approach of Ref. \cite{tcakaev2023intermixing}, we repeated the complete sum-rule analysis 20,000 times while randomly varying the relevant analysis parameters around the values used in the nominal analysis. The varied analysis parameters include the Fe and Nd background-fitting and edge-jump ranges, step-function widths and edge positions, the Fe $L_3$:$L_2$ step ratio, the Nd $M_5$:$M_4$ step ratios, and the first-edge integration cutoff. In addition, the effect of point-to-point experimental noise was included in the uncertainty analysis. The corresponding $M_\mathrm{XMCD}$ distributions are shown in Fig. \ref{Fig:Uncertainty}. We take the standard deviation of each distribution as the estimated analysis uncertainty. The standard deviations are 0.087 $\mu_{\mathrm{B}}$/f.u. at 10 K and 0.086 $\mu_{\mathrm{B}}$/f.u. at 100 K, giving estimated analysis uncertainties of 0.09 $\mu_{\mathrm{B}}$/f.u. at both temperatures. These values characterize the sensitivity of the derived $M_\mathrm{XMCD}$ to the analysis procedure and experimental noise. Table \ref{tab:S1} compares these values with the SQUID magnetization for $H$ $\parallel$ $c$ and, for reference, with the corresponding Nd 4$f$ contributions 
(the $z$-component of the $L \cdot S$-atomic value) and Fe 3$d$ contributions obtained from the collinear DFT calculations. This comparison should be regarded as semi-quantitative for several reasons. 

\begin{table*}[t]
\caption{
Comparison of magnetization values per formula unit estimated from SQUID ($M_\mathrm{SQUID}$), XMCD ($M_\mathrm{XMCD}$ = $3m^{\mathrm{Nd}\text{-}4f}_{\mathrm{tot}}$ + $3m^{\mathrm{eff, Fe}\text{-}3d}_{\mathrm{tot}}$, where $m^{\mathrm{Nd}\text{-}4f}_{\mathrm{tot}}$ = $m_s^{{\mathrm{Nd}\text{-}4}f}$+$m_l^{{\mathrm{Nd}\text{-}4}f}$ and $m^{\mathrm{eff, Fe}\text{-}3d}_{\mathrm{tot}} = m_s^{{\mathrm{eff,Fe}\text{-}3}d}+m_l^{{\mathrm{Fe}\text{-}3}d}$), and DFT. $M_\mathrm{XMCD}$ represents the $c$-axis-projected moment and $m_s^{{\mathrm{eff,Fe}\text{-}3}d} = m_s^{{\mathrm{Fe}\text{-}3}d}-7\langle T_z\rangle$. For comparison, the DFT value includes the corresponding Nd 4$f$  
(the $z$-component of the $L \cdot S$-atomic value) and Fe 3$d$ total moments obtained for the collinear magnetic configuration listed in Table I of the main text. 
}
\label{tab:S1}
\begin{ruledtabular}
\begin{tabular}{lcccc}
Technique & $T$ (K) & $\mu_0H$ (T) & $M$ ($\mu_{\mathrm{B}}$/f.u.) &  \\
\hline
$M_\mathrm{SQUID}$ & 10 & 7 & 0.92 & $H \parallel c$ \\
$M_\mathrm{XMCD}$ & 10 & 9 & $1.35 \pm 0.09$ &
$H \parallel c$ \\
$M_\mathrm{SQUID}$ & 100 & 7 & 2.8 & $H \parallel c$ \\
$M_\mathrm{XMCD}$ & 100 & 9 & $3.75 \pm 0.09$ &
$H \parallel c$ \\
DFT & 0 & -- & 2.10 &
Nd 4$f$ + Fe 3$d$, collinear model \\
\end{tabular}
\end{ruledtabular}
\end{table*}

First, XMCD measures moments projected along the x-ray propagation direction; in the present geometry, this direction is parallel to both the applied magnetic field and the crystallographic $c$ axis. Even for the moments parallel to the x-ray beam, a direct comparison with DFT local moments (collinear configuration) is impeded by the employed open-core approximation. The Nd 4$f$ moments are set to the maximum possible value for the ionic state while the crystal-field effects have been neglected. 

Second, SQUID measures the total bulk magnetization, whereas $M_\mathrm{XMCD}$ includes only the Nd 4$f$ and Fe 3$d$ contributions. The full DFT local-moment decomposition, including the Nd 5$d$ and Sb 5$p$ induced moments, is given in Table I of the main text; these additional contributions are not included in $M_\mathrm{XMCD}$. 

Third, the Fe XMCD value includes the magnetic dipole term, while the Nd spin sum-rule analysis relies on the free-ion Nd$^{3+}$ value of $\langle T_z\rangle/\langle S_z\rangle$; both points are discussed above. The DFT values are for a collinear zero-temperature configuration. Moreover, the SQUID and XMCD values listed in Table \ref{tab:S1} were measured at 7 and 9 T, respectively, and the magnetization remains field dependent in this range. This field mismatch, together with the different quantities probed by the two techniques as discussed above, limits a direct quantitative comparison between the SQUID and XMCD estimates.

\section{Density-functional-theory calculations}

All calculations were performed within the Perdew–Burke–Ernzerhof (PBE) implementation of the generalized gradient approximation (GGA) functional \cite{PBE1996} using the FPLO code \cite{Koepernik1999}, version 22. For Nd $4f$ states, we used the open-core (OC) approach. Within this approximation, the $4f$ states, although partially occupied, are treated as core states. It is also possible to assign a desired spin moment by carefully choosing the up- and down-spin populations. However, due to the spherical symmetry of the core states, orbital moments are not taken into account. Therefore, the OC approach should, in principle, be supplemented with crystal-field calculations of the lowest $J$ multiplet. Since the crystal-field  splittings for Nd$_3$Fe$_3$Sb$_7$ are not available, it is pragmatic to consider the maximum value of the orbital magnetic moments as allowed by Hund’s rules. Here, we consider magnetic Nd$^{3+}$ ions, with the spin moment on the $4f$ shell fixed at $\mu_{\rm S} = -2.45\, \mu_{\rm B}$, as allowed by Hund’s rules for the $4f^{3}$ configuration. The corresponding orbital
moment is $\mu_{\rm L} = +5.72\, \mu_{\rm B}$, resulting in a total $z$-projected magnetic moment of $\mu_{\rm J} = g_J J^z \mu_{\rm B} = 3.27\, \mu_{\rm B}$. 

We used a $k$-mesh with $8 \times 8 \times 24$ divisions. Calculations were performed in the scalar-relativistic approximation. We also carried out constrained fixed-spin-moment calculations to map the energy–magnetization landscape. All calculations were restricted to collinear magnetic configurations.

\begin{figure}[b]
	\begin{center}
		\includegraphics[width=\linewidth]{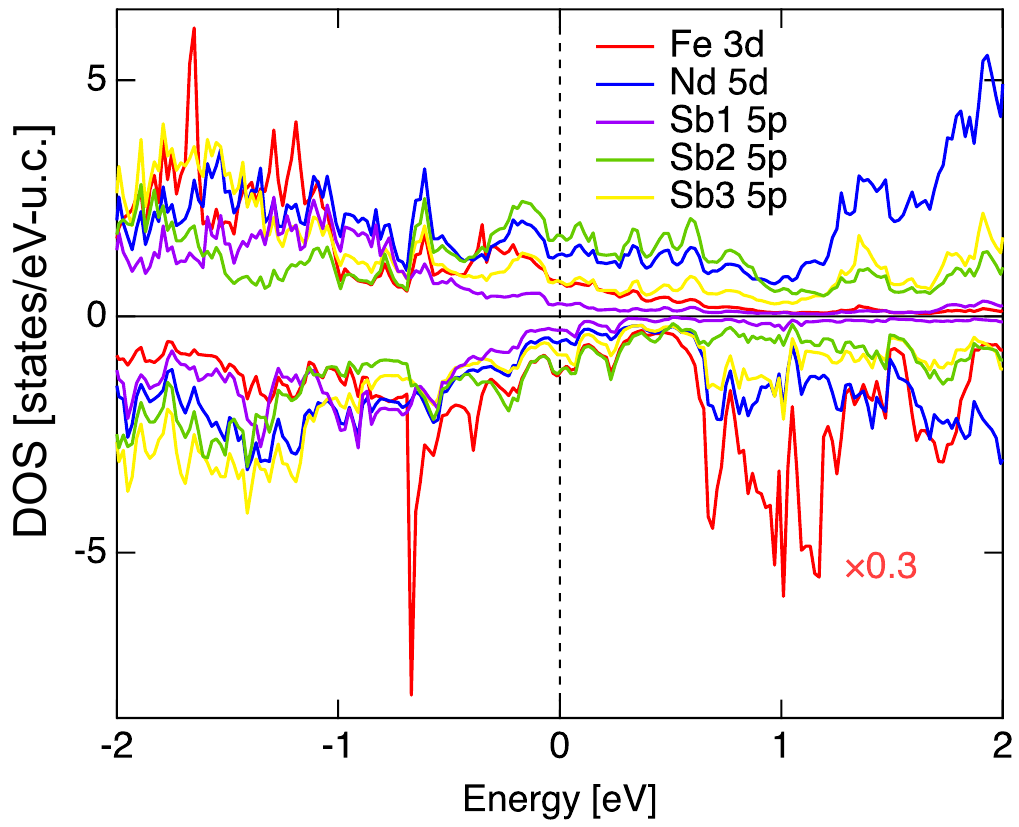}
	\end{center}
	\caption{Expanded view of the DOS near the Fermi energy, showing the projected contributions from Fe 3$d$, Nd 5$d$, and Sb 5$p$ orbitals at the Sb1, Sb2, and Sb3 sites. The Fe 3$d$ DOS is scaled by a factor of 0.3 for clarity.
}
	\label{Fig:DOS}
\end{figure}

To compare candidate magnetic states, we evaluated total energies for a set of collinear configurations denoted by a four-index sign code ($s_1$, $s_2$, $s_3$, $s_4$) with $s_i=\pm$. The indices label the two triangular clusters ($A$,$B$) on each sublattice in order (Nd$_A$, Nd$_B$, Fe$_A$, Fe$_B$); + ($-$) indicates spins parallel (antiparallel) to the $c$ axis. Here $A$ and $B$ refer to the $z=+1/4$ and $z=-1/4$ planes, respectively, for both Nd and Fe sublattices.

We find that the $-+++$ configuration has the lowest total energy across the complete set of collinear four-index configurations (all symmetry-inequivalent sign patterns) evaluated. The DFT results in Table I, including the Nd $5d$ moment in the main text, refer to the $--++$ configuration, which features ferromagnetic order within each sublattice and antiparallel Nd–Fe coupling, and is observed experimentally. The $--++$ configuration is the second-lowest in energy, lying approximately 11 meV per unit cell above the DFT ground state. We attribute the difference between the calculated preference for $-+++$ and the experimentally realized $--++$ state to the collinearity constraint.

\begin{figure}[t]
	\begin{center}
		\includegraphics[width=\linewidth]{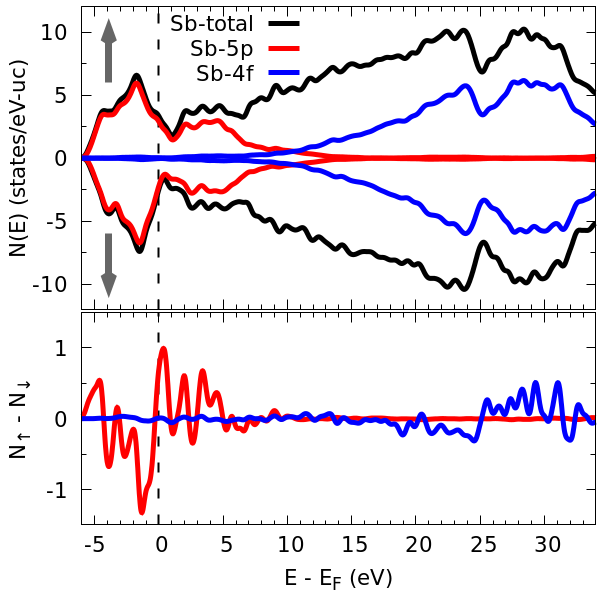}
	\end{center}
	\caption{Orbital-resolved Sb-projected density of states in Nd$_3$Fe$_3$Sb$_7$ calculated by DFT. The total Sb-projected DOS, Sb 5$p$-projected DOS, and Sb 4$f$-projected DOS are shown in black, red, and blue, respectively. The upper panel shows the DOS, and the lower panel shows the corresponding spin-channel differences, $N_\uparrow - N_\downarrow$. The energy is measured relative to the Fermi level $E_\mathrm{F}$.
}
	\label{Fig:DOS_Sb}
\end{figure}

Figure \ref{Fig:DOS} presents an expanded view of the density of states (DOS) near the Fermi energy, showing the contributions from Fe $3d$, Nd $5d$, and Sb $5p$ orbitals at the Sb1, Sb2, and Sb3 sites for the $--++$ configuration. For clarity, the Fe $3d$ DOS is scaled by a factor of 0.3. An energetic overlap between the DOS of spin-polarized Fe $3d$, Sb $5p$, and Nd $5d$ states is observed. 

To clarify the possible final states involved in the Sb $M_{4,5}$-edge spectra, we additionally calculated the orbital-resolved Sb DOS using an extended basis set including Sb 4$f$ final states. In this calculation, Sb 4$s$, 4$p$, and 4$d$ states were treated as semicore states, while Sb 5$s$, 6$s$, 7$s$, 5$d$, 6$d$, 5$p$, 6$p$, 7$p$, and 4$f$ states were included as valence states, following Lejaeghere $et$ $al$. \cite{lejaeghere2016reproducibility}. Figure \ref{Fig:DOS_Sb} shows the orbital-resolved Sb-projected DOS obtained with the extended basis set. The unoccupied states immediately above $E_\mathrm{F}$ are dominated by Sb 5$p$ character, whereas the Sb 4$f$ contribution appears mainly at higher energies and is broadly distributed around 15--20 eV above the Sb 5$p$ band edge. The lower panel shows the corresponding spin-channel differences, $N_\uparrow-N_\downarrow$, which provide a qualitative measure of the spectral weight relevant to the XMCD signal. The Sb 5$p$ spin-channel difference is concentrated near the onset of the unoccupied Sb 5$p$ states, while the Sb 4$f$ contribution is weak near the edge and mainly appears as broad high-energy spectral weight. Therefore, the sharp XMCD structures observed just above the Sb $M_{4,5}$ edges are primarily assigned to Sb 3$d \rightarrow$ Sb 5$p$ transitions.

Although the calculated unoccupied Sb 5$p$ DOS extends from the Fermi level to approximately 10 eV above $E_\mathrm{F}$, this energy extension should not be directly compared with the apparent width of the Sb $M_{4,5}$-edge XMCD peaks. The XMCD intensity reflects the spin-dependent difference in the transition probability, including the energy dependence of the dipole matrix elements and the spin polarization of the final states, rather than the total Sb 5$p$ DOS itself. In the present calculations, the Sb 5$p$ states immediately above $E_\mathrm{F}$ carry the relevant spin polarization associated with Fe 3$d$–Sb 5$p$ hybridization, whereas the broader higher-energy Sb 5$p$ spectral weight contributes much less to the net XMCD signal and may partially cancel between spin channels. The peak widths are also comparable to those reported in previous Sb $M_{4,5}$-edge XMCD studies \cite{ye2015carrier,duffy2018imposing}.

In the configuration $--++$, as already noted in Table I of the main text, the density-functional calculations find that the spin polarization of the Sb $5p$ states is antiparallel to that of Fe $3d$ and Nd $5d$, consistent with the XMCD measurements. This two-fold antiparallel hybridization pathway ($3d-5p-5d$) results in an unexpected antiparallel alignment between the Fe 3$d$ and the total Nd $4f$ moments. Consequently, the Sb atoms are not magnetically passive but instead contribute to intersublattice exchange. In particular, the Sb2 site appears to be most strongly spin polarized and structurally positioned to facilitate indirect coupling between Fe and Nd.

\bibliographystyle{apsrev4-2}
\bibliography{ref}